\documentclass[traditabstract]{aa}

\usepackage{graphicx}
\usepackage{txfonts}
\usepackage{natbib}
\usepackage{longtable,lscape}
\usepackage{epsfig}
\usepackage{fleqn}

\bibpunct{(}{)}{;}{a}{}{,}

\newcommand{\feh}{\mathrm{[Fe/H]}}
\newcommand{\mh}{\mathrm{[M/H]}}
\newcommand{\aFe}{[\alpha/\rm{Fe}]}
\newcommand{\alfe}{\alpha\rm{Fe}}
\newcommand{\teff}{T_\mathrm{eff}}
\newcommand{\logg}{\log g}

\newcommand{\kms}{\rm{km}/\rm{s}}
\def\dex{\rm{dex}}

\def\Gyr{\rm{Gyr}}
\def\figref#1{Fig.~\ref{#1}}

\begin{document}

\title{New constraints on the chemical evolution of the solar neighbourhood and Galactic disc(s)}
\subtitle{Improved astrophysical parameters for the Geneva-Copenhagen Survey}

\titlerunning{Improved GCS astrophysical parameters}
\authorrunning{Casagrande et al.}

\author{L. Casagrande   \inst{1} \and
        R. Sch\"onrich  \inst{1} \and
	M. Asplund      \inst{1} \and
        S. Cassisi      \inst{2} \and
        I. Ram\'{\i}rez \inst{1,3} \and
	J. Mel\'endez   \inst{4} \and
        T. Bensby       \inst{5} \and
        S. Feltzing     \inst{6}
       }

\institute{Max Planck Institute for Astrophysics,
           Postfach 1317, 85741 Garching, Germany \and
           INAF-Osservatorio Astronomico di Collurania, via Maggini, 
           64100 Teramo, Italy \and 
           The Observatories of the Carnegie Institution for
           Science, 813 Santa Barbara Street, Pasadena, CA 91101, USA\and
           Departamento de Astronomia do IAG/USP, Universidade de S\~ao 
           Paulo, Rua do Mat\~ao 1226, S\~ao Paulo, 05508-900, SP, Brasil \and 
	   European Southern Observatory, Alonso de Cordova 3107, Vitacura,
           Casilla 19001, Santiago 19, Chile \and
           Lund Observatory, Box 43, 22100 Lund, Sweden
          }

\date{Received; accepted}

\abstract{
We present a re-analysis of the Geneva-Copenhagen survey, which benefits from the infrared flux method to improve the accuracy of the derived stellar effective temperatures and uses the latter to build a consistent and improved metallicity scale. Metallicities are calibrated on high-resolution spectroscopy and checked against four open clusters and a moving group, showing excellent consistency. The new temperature and metallicity scales provide a better match to theoretical isochrones, which are used for a Bayesian analysis of stellar ages. With respect to previous analyses, our stars are on average $100$~K hotter and $0.1$~dex more metal rich, which shift the peak of the metallicity distribution function around the solar value. From Str\"omgren photometry we are able to derive for the first time a proxy for $\aFe$ abundances, which enables us to perform a tentative dissection of the chemical thin and thick disc. We find evidence for the latter being composed of an old, mildly but systematically alpha-enhanced population that extends to super solar metallicities, in agreement with spectroscopic studies. Our revision offers the largest existing kinematically unbiased sample of the solar neighbourhood that contains full information on kinematics, metallicities, and ages and thus provides better constraints on the physical processes relevant in the build-up of the Milky Way disc, enabling a better understanding of the Sun in a Galactic context.}


\maketitle

\section{Introduction}

Late-type dwarf stars are long-lived objects and can be regarded as snapshots 
of the stellar populations that are formed at different times and places over 
the history of our Galaxy. Not only their kinematics carry residual information 
on their dynamical histories, but their atmospheres retain a fossil record of 
the composition of elements in the interstellar medium at the time and place of 
their formation. Therefore, F, G, and --to a lesser extent-- K dwarfs have been 
traditionally used to study various aspects of the chemical evolution of the 
Milky Way.

The region in the Milky Way for which this task can be most easily
achieved is the solar neighbourhood; starting from pioneering works
using spectra or ultraviolet and colour excess to estimate the metal
abundance of stars in a Galactic context
\citep[e.g.,][]{wallerstein62,vandenBergh62,els62,schmidt63}, this
endeavour has continued over the years with steadily improving
spectroscopic and photometric studies. The latter
\citep[e.g.,][]{twarog80,olsen83,stromgren87,nordstrom04,haywood08}
comprise large catalogues, but have to pay for this by being only able
to derive one single parameter for metallicity, and no detailed
elemental abundances. On the other hand, spectroscopic studies are
still limited to small samples of a few hundred or about a thousand
stars at most. While some studies \citep[e.g.,][]{edvardsson93,
  favata97, fuhrmann08} rely on kinematically unbiased samples, many
investigations
\citep[e.g.,][]{feltzing98,bensby03,reddy06,bensby07,ramirez07,soubiran08} make
use of sophisticated kinematic selections to achieve significant
numbers of members belonging to different subpopulations in their
sample. Even though the abundance trends in these studies are
better traced thanks to this strategy, a quantitative interpretation
can be more difficult.

Galactic chemo-dynamic studies are now entering a new realm with current 
(e.g.,~RAVE \citealt{steinmetz06}; SDSS \citealt{ivezic08}) and forthcoming 
(e.g.,\ SkyMapper, APOGEE, HERMES, LSST, Gaia) large photometric, 
spectroscopic and astrometric surveys targeting different and fainter 
components of the Galaxy.
These tremendous observational efforts, however, must be supported by equal 
investments to minimize the errors that plague the determination 
of stellar parameters. The most important parameter is the effective 
temperature ($\teff$): its determination has implications for the derived 
abundances, for surface gravities and for the inferred ages, masses, and 
distances via isochrone fitting. If we aim to deconstruct the formation and 
evolution of 
the Milky Way in a star-by-star fashion, it is fundamental to have full 
control over all potential sources of errors.

The preferred stellar $\teff$ scale has been a long debated issue, with various scales differing systematically by $100$~K or more. Though this is still true in many areas of the HR diagram, recent data on solar twins \citep{melendez09:twins,ramirez09}, new data and analyses of interferometric angular diameters \citep{boy10,chiava10} and improved HST absolute spectrophotometry \citep{bohlin07} have allowed to pin down the source of these discrepancies via the infrared flux method (IRFM). This gives a good base for the zeropoint of the temperature scale of dwarfs and subgiants, which has now an uncertainty of the order of only $20$~K \citep{casagrande10}. For solar-type stars, the new IRFM scale supports effective temperatures approximately $100$~K hotter than those of \cite{alonso96:teff_scale}, which has been the {\it de facto} choice in many studies until now. Such a shift on the zeropoint has an immediate consequence on the abundances and ages derived for nearby, solar-like stars \cite[see also][]{melendez10} and therefore for interpreting the most basic constraints on Galactic chemical evolution models, namely the metallicity distribution function and the age--metallicity relation. The HR diagram constructed using our newly derived $\teff$ scale matches very well that predicted by stellar models for evolved F and G dwarfs \citep{vandenberg10,brasseur10}, thus avoiding the introduction of any {\it ad hoc} shifts to the $\teff$ scale as was the case in some previous studies \citep[e.g.,][]{nordstrom04}.

The purpose of the present work is to carry out a revision of the 
astrophysical parameters in the Geneva-Copenhagen Survey 
\citep{nordstrom04,holmberg07,holmberg09} with the new effective temperature 
scale presented in \cite{casagrande10} as a starting point to derive new 
metallicities and ages. We improve not only on the accuracy, i.e.~reduce 
zeropoint systematics, but also the precision by reducing internal errors 
stemming from photometric transformations, resulting in highly homogeneous 
astrophysical parameters. These improvements turn out to be crucial to provide 
more stringent observational 
constraints on Galactic chemical evolution theories and hence on the history 
of the Milky Way. In fact, a knowledge of the metallicity distribution together 
with Galactic abundance gradients can improve our understanding of the impact 
and shape of the stellar migration process in the Galactic disc 
\citep[][]{ralph09a,ralph09b}. 
Because models including radial migration relax the classical tight 
correlation between age and metallicity, this relation becomes 
effectively an additional constraint independent from the metallicity 
distribution. 

As we will demonstrate, an estimate of $\aFe$ for most of the stars is also 
obtained here for the first time from Str\"omgren indices.  
Having an indication of $\aFe$ allows for a tentative dissection of the 
chemical thin and thick disc. These estimates are far less 
accurate than those obtained by high-resolution spectroscopy, yet this sample 
exceeds the largest spectroscopic studies available so far by more than an 
order of magnitude and it is not biased by any kinematic selection.

The paper is organized as follows. We present the sample and the determination 
of new effective temperatures and metallicities in Section \ref{sec:intro}. 
Correspondingly, new ages and masses for the stars are derived in Section 
\ref{sec:age}. 
In Section \ref{sec:mdf} we use this information for studying the metallicity 
distribution function in the solar neighbourhood and briefly discuss a possible 
signature of the Galactic bar. The age--dispersion relation is discussed in 
Section \ref{sec:adr}, while Section \ref{sec:disc} is devoted to a better 
understanding of the disc and its metallicity gradient. We finally present 
our conclusions in Section \ref{conclusions}.

\section{Determination of astrophysical parameters}\label{sec:intro}

The Geneva-Copenhagen Survey (GCS) is the most comprehensive catalogue of 
late-type solar neighbourhood stars, providing kinematics and Galactic orbits 
for a 
magnitude-limited and kinematically unbiased sample of $16682$ of 
FG(K) dwarfs brighter than $V\sim8.3$. Some $63000$ radial velocity 
measurements were used to assemble 
the catalogue, which, complemented with Tycho2 proper motions and {\it 
Hipparcos} parallaxes, also provides binarity indication and distances. Because 
the selection of stars into the 
final catalogue was purely based on colour and magnitude cuts, the survey 
provides (apart from effects by the photometric selection) a kinematically 
unbiased census of the solar neighbourhood. While we do not have access to the 
original sample selection performed in assembling the catalogue, we refer to 
\cite{nordstrom04} for a comprehensive discussion on the adopted selection 
criteria and relative completeness. 
Homogeneous Str\"omgren 
photometry was used to derive $\teff$ and $\feh$ for nearly all stars in the 
survey. The original catalogue \citep[][GCSI]{nordstrom04} has undergone a 
number of revisions to improve the temperature and metallicity calibrations 
\citep[][GCSII]{holmberg07} and to account for the new reduction of the 
{\it Hipparcos} parallaxes \citep[][GCSIII]{holmberg09}. 
\begin{figure}
\includegraphics[scale=0.7]{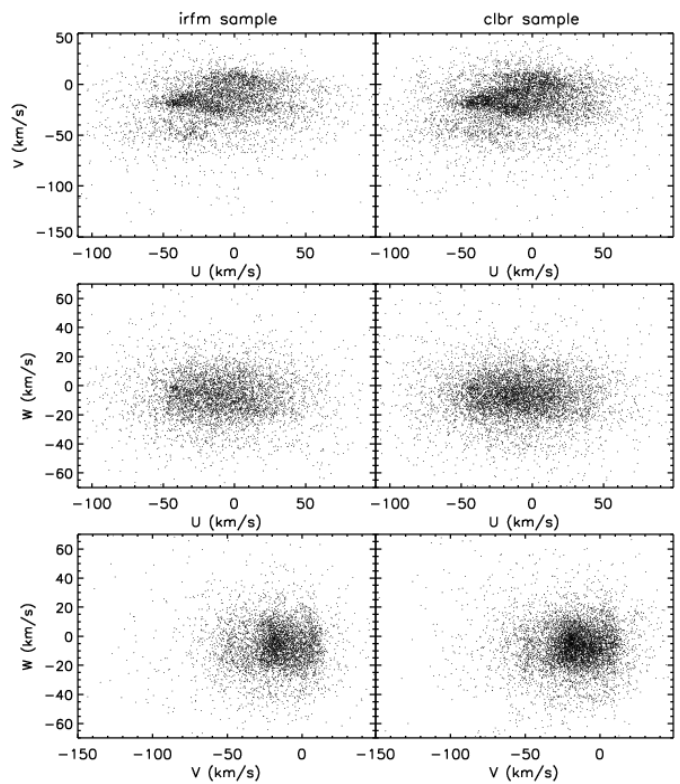}
\caption{Distributions in the velocity planes of the two subsamples defined 
in this work.}
\label{f:uvw_plane}
\end{figure}

However, recent work has shown that the temperature scale adopted in GCSI-III 
is too cold \citep{casagrande10,melendez10}. This has 
far-reaching implications: hotter temperatures imply higher spectroscopic 
metallicities and --when relying on stellar isochrones-- lower age estimates. 
\begin{figure*}
\includegraphics[scale=0.71]{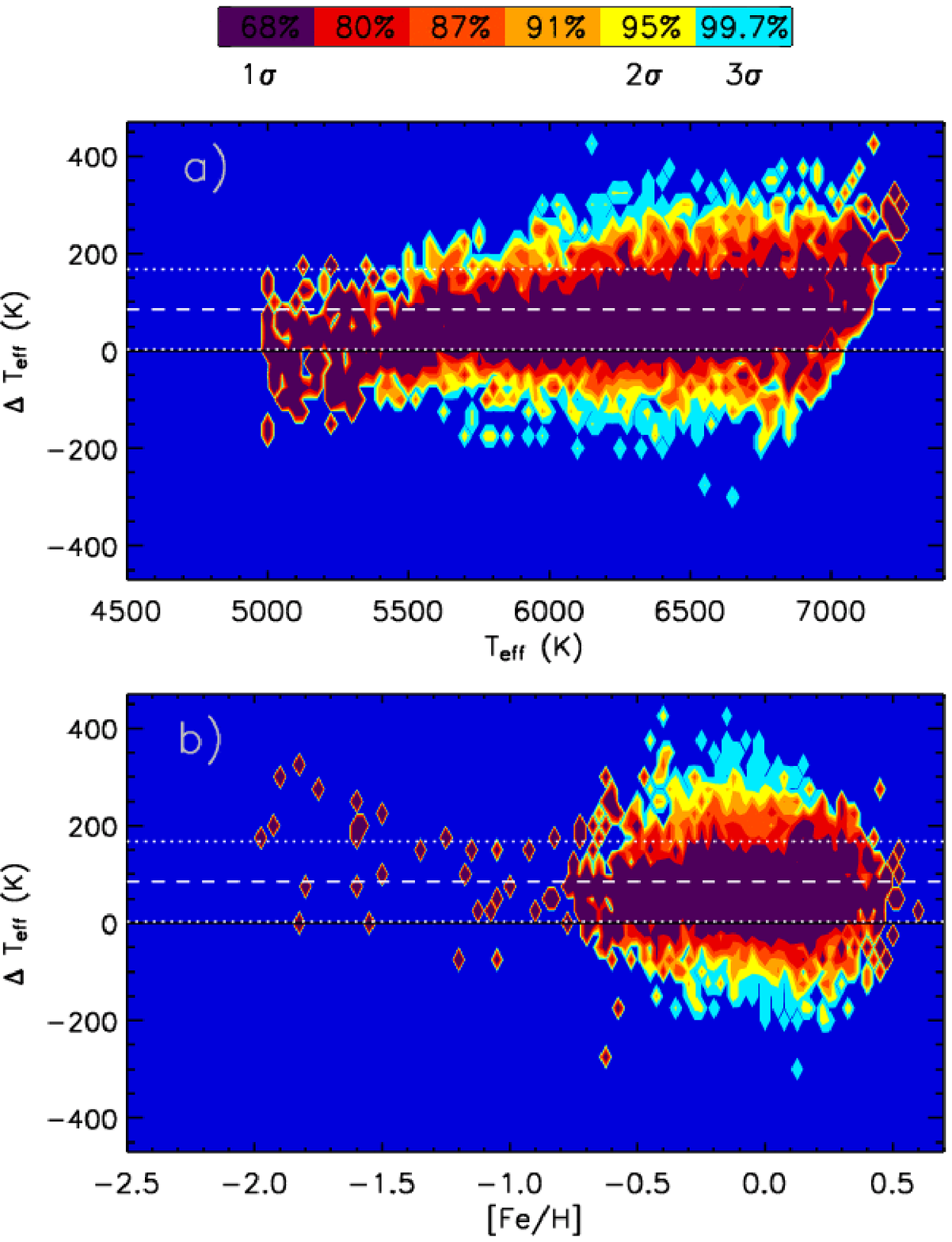}
\includegraphics[scale=0.71]{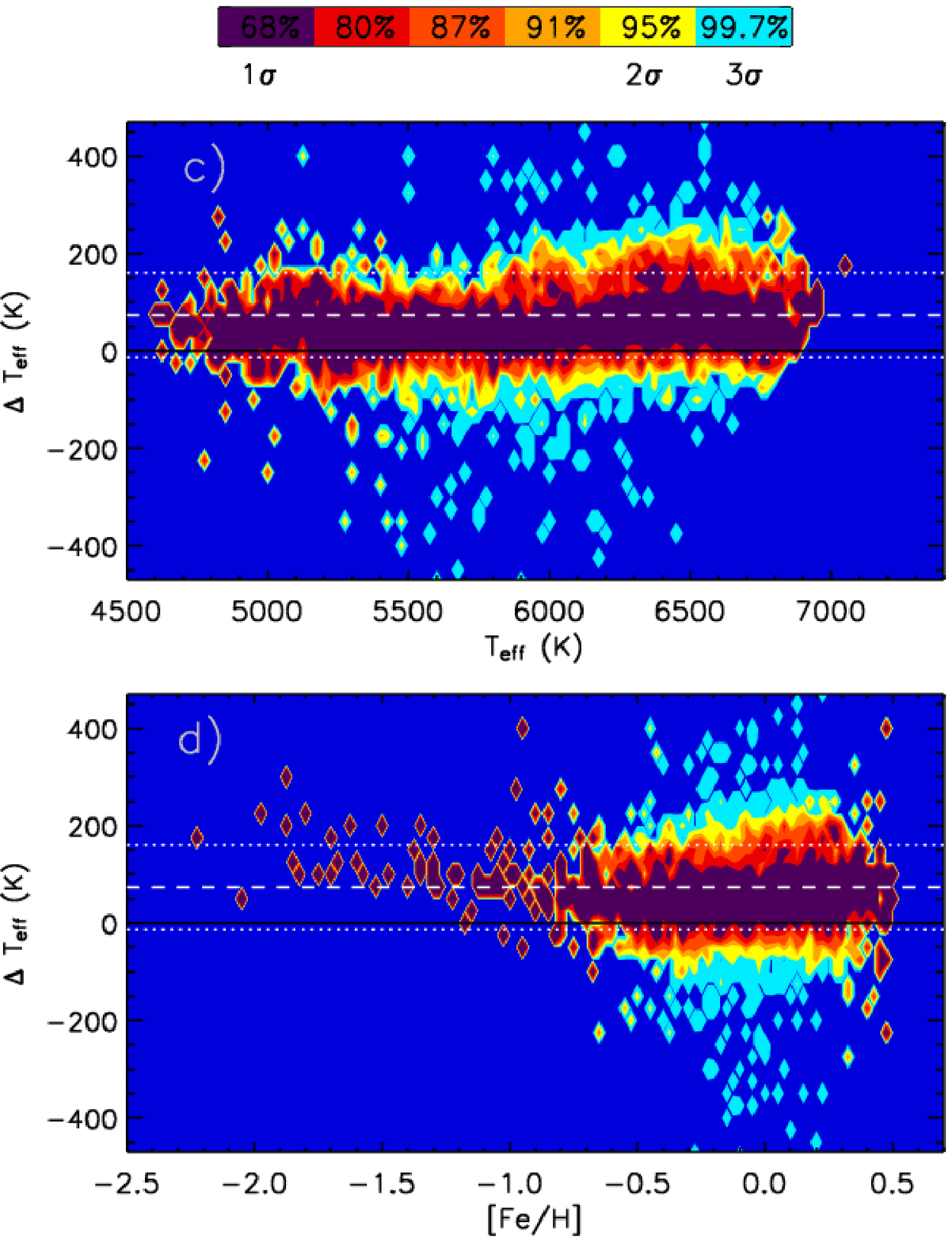}
\caption{Panel a) and b): $\Delta\teff$ (ours minus GCSII) as a 
function of our $\teff$ (upper) and $\feh$ (lower) for stars in the 
{\it irfm} sample.
Panels c) and d): same as before, but for all remaining 
GCS stars in the {\it clbr} sample (see discussion in Section \ref{sec:teff}). 
Contour levels are computed on abscissa intervals of 25~K and 
$0.025$~dex, respectively, to equally represent regions with fewer stars. 
Dashed and dotted lines indicate the mean difference and standard deviation of 
the entire sample; because they are symmetric and are dominated by regions with 
the highest overdensity of stars, the dashed and dotted lines are in some cases 
offset from the local $1 \sigma$ contour levels.}
\label{f:teff}
\end{figure*}

Because we use photometry to derive astrophysical parameters, it is
crucial to clean the sample from binaries, variable stars and/or less
certain measurements.  A description of our selection leading to stars
with the best photometry ({\it irfm} sample) with respect to the
remaining ones ({\it clbr} sample) is given in Section \ref{sec:teff},
where we also briefly present our implementation of the IRFM and the
new effective temperatures derived for the entire GCS catalogue.
Notice that the distinction between the two samples is based
exclusively on the photometric criteria applied and therefore does
not introduce any apparent kinematic bias (Fig.~\ref{f:uvw_plane}).
The corresponding new metallicity scale and ages are then discussed in
Section \ref{sec:feh} and \ref{sec:age}, respectively.

\subsection{A new effective temperature scale}\label{sec:teff}

The effective temperatures in the GCSI were derived using the Str\"omgren 
calibration of \cite{alonso96:teff_scale}, which however lacked a 
sufficient number of stars bluer (i.e.~hotter) than $(b-y) \sim 0.3$ 
($\teff \sim 6500$~K). 
In GCSII this problem was tackled by deriving a new $(b-y)$ vs.\ $\teff$ 
calibration, where effective temperatures for all stars were first obtained 
using the $(V-K)$ calibration of \cite{diBene98} and then $\teff$ were 
re-derived by applying this new $(b-y)$ calibration to the full catalogue. 
However, the calibration of \cite{diBene98} is defined in Johnson $K$, 
enforcing a colour transformation from the 2MASS $K_S$ used in GCSII. Because 
the standards of the Johnson system are all saturated in 2MASS, this renders 
the transformation between the two systems less precise \citep{carpenter01}.
In addition, the metallicity effect is largely reduced but not zero even in 
$(V-K)$, and the calibration of \cite{diBene98} does not account for this 
effect.
The zeropoint of the \cite{diBene98} temperature scale is roughly intermediate 
between that of the \cite{alonso96:teff_scale} scale and our own, which is some 
$50$~K hotter \citep[see below and the comparison in][]{casagrande06}. 

\subsubsection{Reddening}\label{sec:red}

When deriving $\teff$ from photometry it is crucial to correct for 
reddening, if present. Fortunately, given the solar neighbourhood nature of 
the sample used here, most of the stars are unaffected by this problem. 
Reddening estimates derived from Str\"omgren photometry are known to be 
generally reliable (in this case with a stated precision $\sigma_{E(b-y)} 
= 0.009$ mag, \citealt{holmberg07}, but see also \citealt{ks10} for a recent 
revision), and we adopted a procedure similar to GCS for all 
stars\footnote{In the {\it irfm} sample the 
colour excess has been scaled according to intrinsic colour of the star 
\citep[see][]{casagrande10} from which the following mean 
reddening relations were computed and used for the {\it clbr} sample:
$E(B_T-V_T)=1.32\,E(b-y)$, $E(V_T-J)=3.18\,E(b-y)$, $E(V_T-H)=3.66\,E(b-y)$, 
$E(V_T-K_S)=3.93\,E(b-y)$ and $R_{B_{T}}=4.23$, $R_{V_{T}}=3.24$, $R_J=0.86$, 
$R_H=0.50$, $R_{K_{S}}=0.30$, where $R_\zeta=\frac{A(\zeta)}{E(B-V)}$, 
$A(\zeta)$ is the extinction in a given $\zeta$ band.
For the Str\"omgren indices $E(m_1)=-0.30\,E(b-y)$ and $E(c_1)=0.20\,E(b-y)$ 
were adopted from \cite{cb70}.} i.e.~a reddening correction is applied 
for stars with $E(b-y)$ greater than $0.01$ mag and 
farther away than 40~pc, otherwise a value of zero is assumed. 

Thus, only about one quarter of the stars in the GCS catalogue need to be 
corrected for reddening, the median $E(b-y)$ being $0.02$~mag, as one would 
expect given the nearby nature 
of the sample \citep[see also][for a plot of the colour 
excess in different distance intervals]{holmberg07}. Note that the effect of 
colour excess on $\teff$ derived via IRFM amounts to $\sim 50$~K for every 
$0.01$~mag \citep[see][for further details]{casagrande10} whereas in the case 
of colour-temperature calibrations its effect can be directly estimated.

Although the temperature and metallicity scales are unchanged between GCSII and 
III, we noticed important differences between the two catalogues. These 
differences show a correlation with the adopted $E(b-y)$, reaching several 
hundred K in $\teff$ and almost $1\,\dex$ in $\feh$ for stars with the highest 
colour excesses. This suggests that stars in GCSIII have not been corrected 
for reddening and because of this we will only use the kinematic data from 
GCSIII and the stellar parameters from GCSII when making comparisons.
\begin{figure}
\includegraphics[scale=0.71]{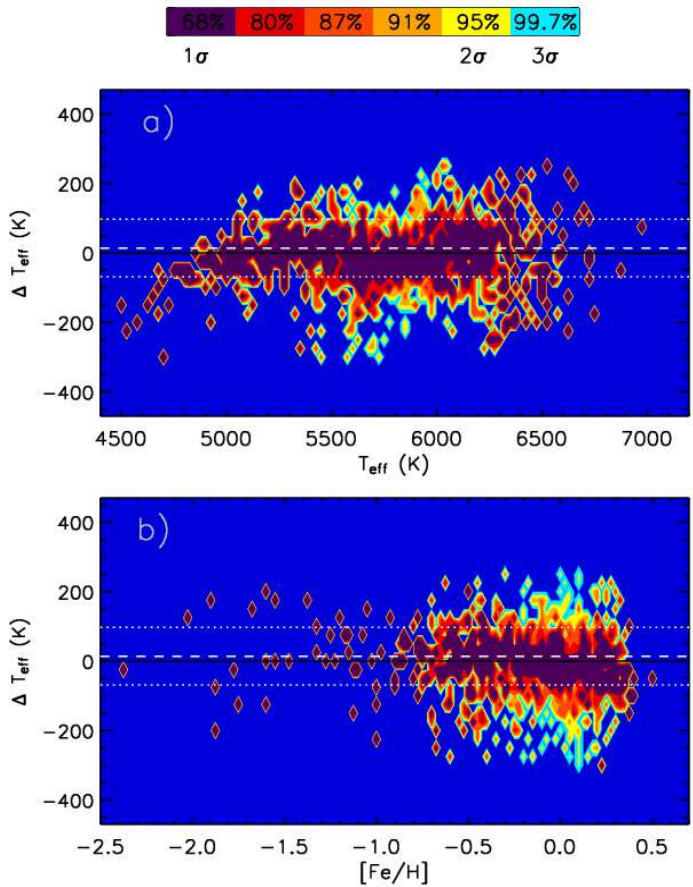}
\caption{$\Delta \teff$ (ours minus spectroscopic values) for the 1498 stars 
in our calibration sample as function of $\teff$ (panel a) and $\feh$ 
(panel b). Contour levels and lines as in Fig.~\ref{f:teff}.}
\label{f:teff_spec}
\end{figure}

\subsubsection{{\it irfm} sample}\label{sec:irfm}

The IRFM implementation described in \cite{casagrande10} not only improves 
the accuracy of the zeropoint of the derived stellar parameters, but 
also their precision by employing Tycho2 $B_TV_T$ and 2MASS $JHK_S$ 
photometry to simultaneously recover the bolometric flux 
--$\mathcal{F}_{Bol}(\rm{Earth})$-- and the effective temperature of each star. 
It is well suited to be applied to the Geneva-Copenhagen catalogue 
directly, avoiding the use of colour calibrations as well as transformations 
among different photometric systems. 

Given its nature, it is crucial to have good photometry in all bands for the 
stars we apply the IRFM to. From the GCSII we exclude stars flagged as variable 
or having multiple components. We retrieve Tycho2 $B_TV_T$ magnitudes for all 
targets \citep{hog2000} and additionally cross-check and discard those 
classified as variable or non-single in {\it Hipparcos}. 
The faintest stars might have uncertain photometry in Tycho2, whereas the 
brightest can be saturated in 2MASS: when applying the IRFM we consider only 
stars with photometric errors $\sigma_{B_{T}} + \sigma_{V_{T}} < 0.10$ and 
``j\_''$+$``h\_''$+$``k\_msigcom''$<0.10$ all having quality flag ``A'' in 
2MASS\footnote{i.e. with best photometric detection http://www.ipac.caltech.edu/\\
2mass/releases/allsky/doc/sec1\_6b.html\#phqual}. Stars having $\teff < 5000$~K emit considerable amount of flux in the 
red. The computation of the bolometric flux (and thus $\teff$) is inaccurate 
if one uses only Tycho2 and 2MASS photometry \citep{casagrande10} and 
therefore we exclude stars cooler than this limit. This cut concerns 
only a minor part ($326$ stars out of $16682$) of the sample in the GCS.

In our implementation of the IRFM an iterative procedure is adopted to cope 
with the mildly model-dependent nature of the bolometric correction: given an 
initial estimate of $\teff$, we interpolate over a grid of synthetic stellar 
fluxes at the appropriate $\feh$ (as determined in Section \ref{sec:feh}) and 
$\logg$ of each star, until convergence in $\teff$ is reached within $1$~K. 

For all stars, $\logg$ is determined from the fundamental relation
\begin{equation}\label{eq:logg}
\log \frac{g}{g_{\odot}} = \log \frac{\mathcal{M}}{\mathcal{M}_{\odot}} + 4 \log \frac{\teff}{T_{\odot}} - \log \frac{L}{L_{\odot}},
\end{equation}
where $L$ is the bolometric luminosity\footnote{In this work 
we take $L_{\odot}=3.842 \times 10^{33} \rm{erg s^{-1}}$ \citep{bahcall06}.} and 
$\mathcal{M}$ is the mass of the star, obtained by interpolating over 
isochrones. Notice that in Eq.~(\ref{eq:logg}) mass plays only a secondary 
role: varying it by 10\% changes $\logg$ by 0.04~dex. We used the masses 
reported in GCSII as a starting value and the BASTI mass expectation 
values (cf.~appendix) for a second iteration.
Even variations as large as 
$\pm 0.5$~dex in surface gravity change the $\teff$ obtained via IRFM by only 
a few tens of a K \citep{casagrande06,casagrande10}, thus having negligible 
impact. The bolometric luminosity $L$ is computed from 
$\mathcal{F}_{Bol}(\rm{Earth})$ using the new 
{\it Hipparcos} parallaxes \citep{vanLeeuwen07}, and an iterative procedure 
was adopted to converge in $\logg$ using at each step the corresponding 
effective temperature and luminosity obtained from the IRFM. Although in the 
GCSI a photometric selection was made to cut out giant stars, there is a 
handful of them left. We exclude those labelled as suspected giants in the GCS 
and restrict the {\it irfm} sample to $\logg \ge 3.0$.
Altogether, we are left with a sample of 6670 stars that satisfy all of the 
above criteria on photometric quality, non-binarity, and surface gravity. To 
these we can apply the IRFM. A MonteCarlo simulation using 
the measured observational errors ($\sigma_{B_{T}}, \sigma_{V_{T}}$, 
``j\_'', ``h\_'', ``k\_msigcom'' and $\sigma_{\feh}$) was used to estimate 
the random error in the resulting $\teff$ and $\mathcal{F}_{Bol}(\rm{Earth})$ 
of each star, to which the systematic uncertainty arising from the adopted 
absolute calibration was added \citep[see][]{casagrande06,casagrande10}. 

\subsubsection{{\it clbr} sample}\label{sec:clbr}

For all remaining stars in the GCSII effective temperatures and 
bolometric fluxes were computed using the colour calibrations in $(b-y)$, 
$(B_T-V_T)$, $(V_T-J)$, $(V_T-H)$ and $(V_T-K_S)$ from \cite{casagrande10}, 
which extend also below $5000$~K. We only took into consideration photometry 
with $\sigma_{B_{T}} < 0.05$, 
$\sigma_{V_{T}} < 0.05$, ``j\_msigcom''$<0.04$, ``h\_msigcom''$<0.04$, 
``k\_msigcom''$<0.04$ (which in the colour--temperature relations imply formal 
uncertainties similar to those of the stars analysed using the IRFM). 
We computed the average $\teff$ and $\mathcal{F}_{Bol}(\rm{Earth})$ if more 
than one index was used and applied a $3\sigma$ clipping if more than two 
indices were present. In the latter case, the standard deviation was used to 
estimate the error in the derived $\teff$.
Notice that these calibrations (as any 
available in literature) do not include an explicit $\logg$ dependence. 
However, because surface gravities of dwarfs and subgiants decrease when 
moving to hotter $\teff$, an intrinsic dependence on such a term is likely to 
be built into them \citep[see also the discussion in][]{casagrande10}. Our 
colour--temperature calibrations indeed perform extremely well along most of 
the CMD morphology defined by F and G dwarfs and subgiants \citep{vandenberg10}.

Figure \ref{f:teff} shows the comparison between $\teff$ derived in 
Section \ref{sec:irfm} and \ref{sec:clbr} and those in GCSII. A mean 
difference of about $100$\,K appears, and there are trends at the highest 
and lowest $\teff$, as well as at the lowest metallicities. The latter trend 
could arise from the absence of an explicit metallicity dependence in 
\cite{diBene98} or from the limitation of the standard functional form 
used in literature when fitting effective temperatures and metallicities as 
function of $(b-y)$ \citep[see][]{casagrande10}. We note that the IRFM depends 
only marginally on the assumed metallicity, and we verified for the GCS stars 
that changing $\feh$ by 
$\pm 0.2$~dex affects $\teff$ by about 20\,K at most. The impact can indeed be 
larger when one uses colour--$\teff$ relations that involve optical bands. 
 
\subsection{A new Str\"omgren metallicity scale}\label{sec:feh}

The $uvby$ photometric\footnote{In the following, we will refer to $(b-y)$, 
$m_1$ and $c_1$ with the implicit understanding that they were dereddened 
if there was any colour excess. In the same manner, absolute magnitudes 
were corrected as well when necessary.} 
system \citep{stromgren63} is well suited for the 
determination of basic stellar atmospheric parameters through the colour 
indices $(b-y)$, $m_1 = (v-b)-(b-y)$ and 
$c_1 = (u-v)-(v-b)$. The $m_1$ index is designed to measure the depression 
owing to metal lines around $4100$~\AA~ ($v$ band), and hence is suitable to 
infer the metallicity in a variety of stars \citep[e.g.,][and references 
therein]{bessell05}. The $c_1$ index is 
designed to evaluate the Balmer discontinuity, which is a temperature indicator 
for B- and A-type stars and a surface gravity indicator for late-type stars, 
though for stars comparable to or cooler than the Sun it also carries 
metallicity information \citep[e.g.,][]{twarog02,onehag09,melendez10}. 
Several calibrations exist in the literature that link Str\"omgren colours to 
astrophysical parameters, following either theoretical (i.e. based on model 
atmospheres) or empirical approaches (see \citealt{onehag09} and 
\citealt{arnadottir10} respectively, for recent reviews).
\begin{figure*}
\includegraphics[scale=0.6]{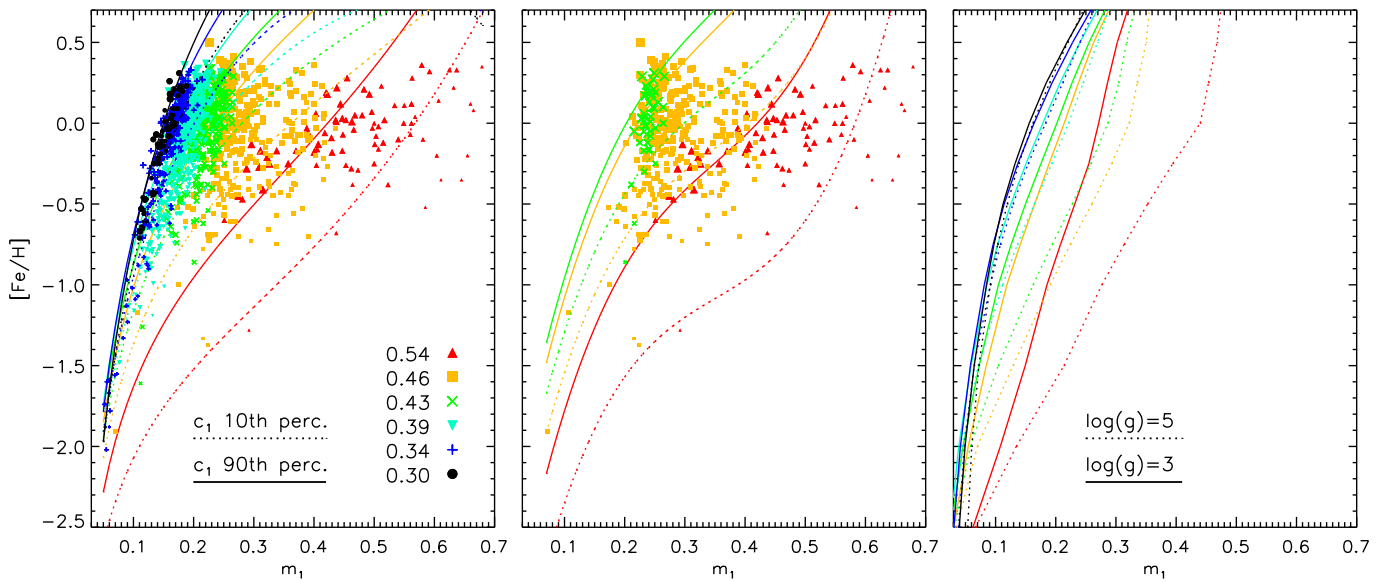}
\caption{Left panel: $\feh$ versus $m_1$ for our 1498 
calibrating stars in different $(b-y)$ ranges represented by 
different symbols (median values as indicated in the labels). The 
size of the symbols increases with higher values of $c_1$.
Dotted and continuous lines represent Eq.~(\ref{eq:m1_feh}) at these 
median values for 
the 10th and 90th percentiles of the $c_1$ distribution. Central 
panel: same as left panel, but for Eq.~(\ref{eq:c1_feh}), which 
applies only to cool stars. 
Right panel: theoretical $\feh$ versus $m_1$ relation when using 
synthetic colours from the ``MARCS-standard'' library and $\logg$ 
instead of $c_1$ (see text for discussion).}
\label{f:m1}
\end{figure*}

  Throughout the paper, we talk of metallicity both in terms of
  iron abundance $\feh$ and
  overall metal content $\mh$, if a clear distinction is not
  needed. Notice though that Eq.~(\ref{eq:m1_feh}) and
  (\ref{eq:c1_feh}) given later in this section are calibrated 
  against spectroscopic measurements
  of $\feh$ and thus are strictly referring to iron abundance. The
  overall metal content --always indicated by $\mh$ in this work-- was 
  obtained using the same functional forms, but accounting for
  $\aFe$ in the calibration sample. Later in this section we develop a
  new estimator for the $\alpha$-element content in most of the GCS 
  stars.

The metallicity calibration adopted in GCSII patches the red ($b-y \ge 0.46$) 
and blue ($b-y \le 0.30$) calibrations derived in 
the GCSI with a new calibration containing all possible combinations of 
$(b-y)$, $m_1$ and $c_1$ to third order for $0.30 < (b-y) < 0.46$. Those three 
calibrations are built by linking Str\"omgren indices to spectroscopic 
metallicities gathered from a large number of studies, affecting the 
homogeneity of the results. In addition, the calibration in the 
$0.30 < (b-y) < 0.46$ range is based on spectroscopic studies with a 
$\teff$ scale broadly consistent with that adopted in the GCSII, i.e.~cooler 
than the one used in this study, implying an offset in the zeropoint of 
the metallicity scale. The adopted $\teff$ scale is in fact the main driver in 
setting the zeropoint of the metallicity scale.

Over the past few years an increasing number of high-resolution and high 
signal-to-noise spectroscopic investigations have targeted hundreds of stars in 
the solar neighbourhood. This allows us to build a large and homogeneous 
spectroscopic catalogue, which we use to derive a new metallicity calibration.  
To this purpose we have taken only three large surveys, namely 
\citet[V05]{valenti05}, \citet[S08]{sousa08} and Bensby et al.~(2011 in prep., 
B11, which includes over 600 stars in addition to 102 from \citealt{bensby03} 
and \citealt{bensby05}). 
Apart from spectroscopically determined $\teff$ and $\feh$, all these surveys 
provide $\alpha$ abundances: Si and Ti 
in the case of \cite{valenti05}, and Mg, Si, Ca, Ti for the other two studies 
\cite[for the Sousa et al.~2008 sample the abundances are given in the 
companion paper of][N09]{neves09}. 
They are all very consistent, with mean differences (all in the sense B11-V05 
and B11-S08 for 142 and 85 stars in common, respectively) of 
$\Delta \feh = 0.034\pm0.004$ ($\sigma=0.050$~dex) and $0.047\pm0.005$ 
($\sigma=0.046$~dex) and $\Delta \aFe = 0.06 \pm 0.01$ ($\sigma=0.16$~dex) 
and $0.01\pm 0.02$ ($\sigma=0.20$~dex). These differences are 
small and consistent with the scatter; we also made an attempt to homogenise 
all stars on a common scale (B11) by 
fitting the differences with respect to B11 as linear or parabolic function 
of $\feh$, $\logg$ and $\teff$ but this approach only had a minor effect on 
the overall metallicity calibration. A comparison with the homogenised 
spectroscopic catalogue of \cite{arnadottir10} confirms this conclusion 
(see below).

Our final sample contains 1522 stars, all having Str\"omgren 
colours, $\feh$ and $\aFe$. If a star was found in more than one study, 
we chose the $\feh$ and $\aFe$ from the one that had $\teff$ 
closest to our estimate. The mean difference between photometric and 
spectroscopic $\teff$ is $13 \pm 95$~K. We also applied a $3\sigma$ clipping 
to remove the major outliers, and obtained a final 
calibration sample of 1498 stars ($\Delta \teff = 14 \pm 83$~K), half of which 
are in the {\it irfm} sample. Fig.~\ref{f:teff_spec} compares our 
effective temperatures with those of the three spectroscopic studies. 
The systematic offset between older photometric and spectroscopic $\teff$ 
\citep[e.g.,][]{ramirez04} is now clearly removed thanks to our new 
IRFM implementation.
There are no significant trends as function of effective temperature, except 
for the very few stars below $\sim 5000$~K where spectroscopic estimates have 
the tendency to return hotter $\teff$ than photometric ones \citep[see also 
the discussion in][]{sousa08}. When plotting $\Delta\teff$ as function of 
$\feh$, the metal-poor stars are on average well reproduced despite an 
increasing scatter. 
There is a minor trend in the range $-0.5<\feh<0.5$~dex, with $1\sigma$ 
contour going from $+50$~K to $-50$~K: this could potentially introduce a mild 
systematic bias (as well as affect the width of the metallicity distribution 
function) of the order of $\mp0.05$~dex throughout this range, though for a 
single star this is below the accuracy of our calibration (see below) and 
spectroscopic measurements are themselves not immune from deficiencies. On 
average there is no significant zeropoint offset or trend. 

In the literature various approaches have been used to calibrate Str\"omgren 
photometry to derive metallicities, either based on how much the colour 
indices $m_1$ and $c_1$ differ from a given standard relation \citep[usually 
derived for the Hyades, e.g.,][]{olsen84,haywood02,vanLeeuwen09} or using 
direct combinations of the Str\"omgren indices $m_1, c_1$ and $(b-y)$. This 
is the choice made in most of the recent works \citep[e.g.,][]{schuster89,haywood02,nordstrom04,ramirez05a,holmberg07,twarog07}.
We adopt the latter approach, but we are aware that even though our 
calibrating sample includes a large number of stars, some regions of the 
$\feh$, $\teff$ and $\logg$ space are less well sampled than others (see also 
Fig.~\ref{f:m1}). To limit possible biases, we checked our findings against 
synthetic colours. Despite the inaccuracies that might still plague synthetic 
Str\"omgren colours \citep[e.g.,][]{melendez10}, in many cases they can be used 
at least to provide guidance on general trends \citep{onehag09}. For this 
work, synthetic indices were computed for the full grid of 
``MARCS-standard''\footnote{http://marcs.astro.uu.se where 
standard refers to the chemical composition, i.e. $\aFe=0.0$ for 
$\feh \ge 0.0$, a linear increase of $\aFe$ from $0.1$ at 
$\feh=-0.25$ to $\aFe=0.4$ at $\feh=-1.0$ and $\aFe=0.4$ for 
$\feh \le -1.0$.} model 
spectra \citep{gustafsson08} using the zeropoints and filter transmission 
curves described in \cite{melendez10}. Note that the purpose of using synthetic 
colours is for verification only, and they do not enter into our calibrations, 
which remain fully empirical. 

Fig.~\ref{f:m1} shows the sensitivity of $m_1$ to $\feh$ for our 1498 
calibrating stars, in different $(b-y)$ (basically $\teff$) ranges. The 
asymptotic 
behaviour towards the most metal-poor stars reflects the decreasing 
sensitivity of $m_1$ in this regime and it can be well represented by 
a logarithmic term \citep{schuster89}. Therefore, rather than including all 
possible combination of indices in some high-order polynomial, we started with 
a simple form of the kind $\log(m_1)+a\,m_1^3$ and introduced mixed terms to 
allow for a change of slope with $(b-y)$ and $c_1$, where the ratio between the 
logarithmic and cubic terms, $a$, was optimized by treating it as a 
free parameter in the fitting process. This accounts for the first six terms 
in the following equation
\begin{displaymath}
\feh = 3.927\,\log(m_1) - 14.459\,{m_1}^3 -5.394\,(b-y)\,\log(m_1) 
\end{displaymath}
\begin{displaymath}
\phantom{\feh = } + 36.069\,(b-y)\,{m_1}^3 + 3.537\,c_1\,\log(m_1)
\end{displaymath}
\begin{displaymath}
\phantom{\feh = } - 3.500\,m_1^3\,c_1 + 11.034\,(b-y) - 22.780\,(b-y)^2
\end{displaymath}
\begin{equation}\label{eq:m1_feh}
\phantom{\feh = }  + 10.684\,c_1 - 6.759\,c_1^2 - 1.548,
\end{equation}
where the additional terms that have a linear and quadratic dependence on 
$(b-y)$ and $c_1$ were introduced after verifying that they improved the 
residuals. We 
also checked that the inclusion of terms of higher order did not lead to any 
further gain. Equation (\ref{eq:m1_feh}) applies to stars in the following 
ranges: $0.23 \le (b-y) \le 0.63$, $0.05 \le m_1 \le 0.68$ and 
$0.13 \le c_1 \le 0.60$ with a standard deviation of $0.10$~dex. We remark 
that for stars with $\feh \lesssim -2$ Str\"omgren indices effectively lose 
sensitivity to metallicity. We verified this using an additional sample of 
26 metal-poor dwarfs taken from \cite{casagrande10} and 
\cite{melendez10:lithium}. Those stars, all in the range 
$-3.3 < \feh < -2.0$, 
did not show any significant dependence on metallicity 
\footnote{This appears not to be the case for very metal-poor 
giants above the horizontal branch where in fact $m_1$ follows 
$\feh$ tightly \citep[e.g.][Ad\'en et al.~to be 
submitted]{aden11}.}
and were therefore not 
used in the fitting process, which was limited to the 1498 stars shown in 
Fig.~\ref{f:m1}. 
\begin{figure}
\includegraphics[scale=0.8]{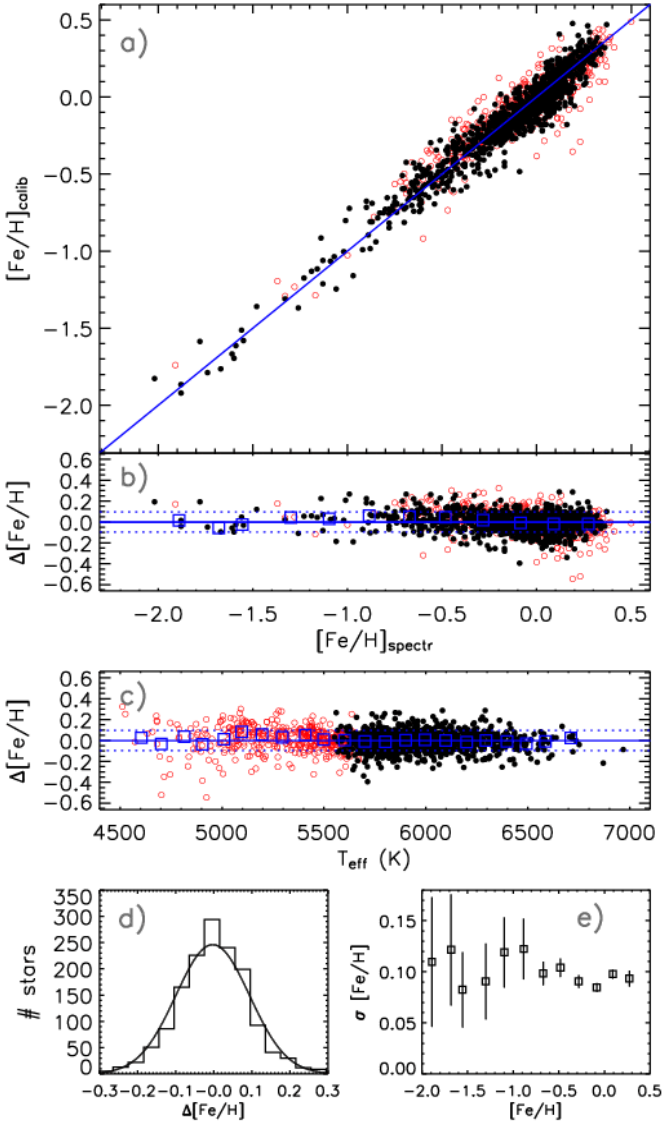}
\caption{Panel a): spectroscopic versus photometric metallicities obtained 
using the calibration presented in Section \ref{sec:feh} for 1498 stars. 
Filled circles are for stars having $(b-y) < 0.43$, open circles for redder 
(i.e.~cooler) stars. Panel b): same as above, but showing residuals 
(ours minus 
spectroscopic). Dotted lines are $1 \sigma$ scatter and boxes the median 
values in non overlapping intervals of $0.2$~dex. Panel c): same as in the 
previous one, but as a function of $\teff$, with boxes computed in non 
overlapping intervals of $100$~K. The overall zeropoint offset is 
$\mu=-0.003$~dex and 
$\sigma=0.097$~dex. Panel d): distribution of the residuals of our calibration 
against spectroscopy with a Gaussian of width $\sigma$ and centred at $\mu$ 
overplotted. 
Panel e): standard deviation associated to each square computed in panel b) 
with error bars being the standard deviation of the mean.} 
\label{f:cal}
\end{figure}
 
While the hottest stars in Fig.~\ref{f:m1} display a remarkably tight 
correlation with $\feh$, for decreasing $\teff$ also $c_1$ correlates well with 
metallicity \citep{twarog02,melendez10}. In fact, different metallicity 
calibrations are often given for F and GK dwarfs separately 
\citep[e.g][]{schuster89,nordstrom04}. While Eq.~(\ref{eq:m1_feh}) applies 
also to cool stars ($(b-y)>0.43$, i.e.~$\teff \lesssim 5600$~K), for those we 
found an additional function of the kind
\begin{displaymath}
\feh = -0.116\,c_1-1.624\,c_1^2 + 8.955\,c_1\,(b-y)
\end{displaymath}
\begin{displaymath}
\phantom{\feh = } + 42.008\,(b-y) - 99.596\,(b-y)^2 + 64.245\,(b-y)^3
\end{displaymath}
\begin{displaymath}
\phantom{\feh = } + 8.928\,c_1\,m_1 + 17.275\,m_1 - 48.106\,m_1^2 
\end{displaymath}
\begin{equation}\label{eq:c1_feh}
\phantom{\feh = } + 45.802\,m_1^3 - 8.467,
\end{equation}
which applies to stars with $0.43 \le (b-y) \le 0.63$, $0.07 \le m_1 \le 
0.68$ and 
$0.16 \le c_1 \le 0.49$ with a standard deviation of $0.12$~dex (the same 
$\sigma$ is obtained considering instead Eq.~\ref{eq:m1_feh} for equally 
red stars, but averaging with this latter form helps to reduce the zeropoint 
offset for cool stars). With respect to the functional form used in 
\cite{schuster89}, ours has the same standard deviation, 
but performs significantly better for $\feh \lesssim -1.0$. 

The right hand panel of Fig.~\ref{f:m1} shows predictions using 
``MARCS-standard'' synthetic colours: models capture the main 
trends, especially at higher $\teff$ and different surface gravities, where 
the choice of various $\logg$ in the synthetic spectra is approximated by the 
$10$ and $90$ percentiles of the $c_1$ distribution in the data (assuming 
lower values of $c_1$ to trace higher $\logg$, which does not hold exactly 
towards the coolest $\teff$, because of contamination between dwarfs 
and subgiants). The main point from Fig.~\ref{f:m1} is that our adopted 
functional form is a good representation of the data, even in poorly sampled 
regions of the plot and the trend at super--solar metallicities, where we do 
have calibration stars, is real.

\begin{figure}
\includegraphics[scale=0.58]{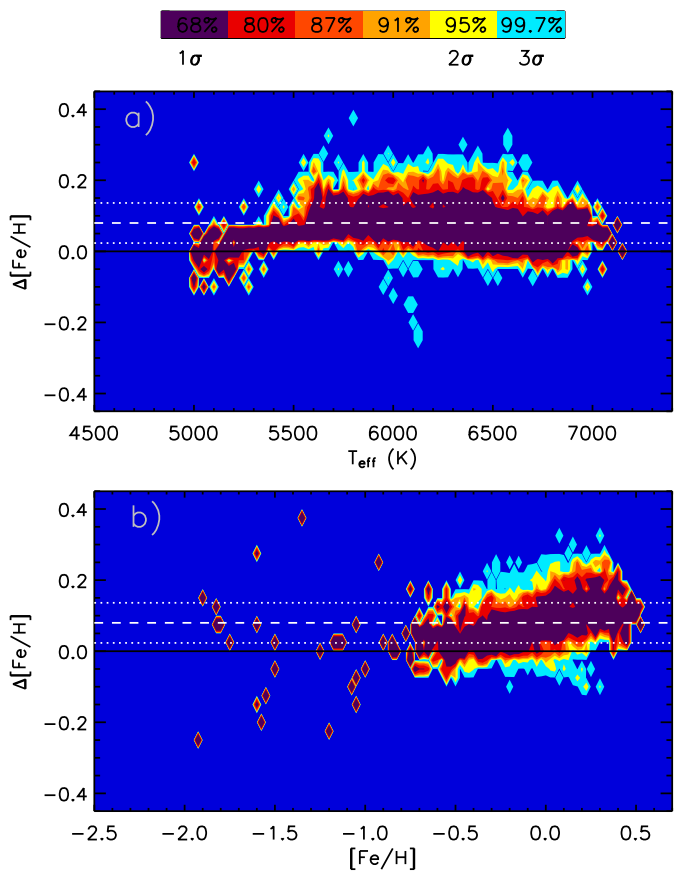}
\caption{$\Delta \feh$ (in the sense ours minus GCSII) as function of $\teff$ 
(panel a) and $\feh$ (panel b) for stars in the {\it irfm} sample 
satisfying the applicability range of our metallicity calibration. Notice the 
two breaks at $5500$~K and $6500$~K which correspond to the discontinuities 
introduced by the three calibrations used in GCSII for different $(b-y)$. 
Contour levels and lines as in Fig.~\ref{f:teff}.}
\label{f:feh}
\end{figure}

We applied Eq.~(\ref{eq:m1_feh}) to all stars in the GCS, but for stars redder 
than $(b-y) \ge 0.43$ we also used Eq.~(\ref{eq:c1_feh}) and then took the 
average of both estimates as our final value. The comparison between the input 
spectroscopic metallicities and our photometrically derived values is shown 
in Fig.~\ref{f:cal}. Both equations provide a good representation of 
spectroscopic measurements and, within their accuracy, we do not introduce any 
obvious discontinuity. Our procedure gives a more homogeneous sample, avoiding 
the presence of breaks in different colour ranges, as was the case in the 
previous GCSII (see Fig.~\ref{f:feh}). 
Uncertainties in the observed Str\"omgren colours also bear on derived 
metallicities. On average, the effect amounts to $0.04-0.05$~dex in 
$\mh$ and $\feh$, estimated running a MonteCarlo 
simulation with observational errors in $(b-y)$, $m_1$ and $c_1$ as 
given in \cite{olsen83}. Errors in the derived metallicities tend to 
increase towards the blue- and red-most indices.

A comparison with the homogenized spectroscopic sample of \cite{arnadottir10} 
confirms the quality of our 
calibration with a median (mean) difference (ours minus Arnadottir) of 
$0.002$ ($0.007$) dex and a scatter $\sigma=0.13$~dex. Note that as discussed 
throughout the text, the overall scatter of our calibration with respect to 
the spectroscopic sample is slightly below $0.1$~dex, though this comparison 
folds the uncertainties that affect spectroscopic estimates. The test 
on open clusters (see Section \ref{sec:further}) suggests that the intrinsic 
scatter in the metal-rich regime is actually somewhat lower.
Using the 
recently determined $uvby$ solar colours of \cite{melendez10}, we obtain 
$\feh_\odot=-0.006$~dex, which agrees well with the zeropoint of 
our metallicity calibration.

\subsubsection{Further test of the [Fe/H] scale}\label{sec:further}

We already checked our metallicity scale against the homogenized 
stellar sample of \cite{arnadottir10} and the solar colours of 
\cite{melendez10}. Here we further test it by using open clusters and a moving 
group; finally we comment upon the limit of our calibration for intrinsically 
bright stars.

Hyades and Coma are two nearby, virtually reddening-free clusters often 
used to check the metallicity scale \citep[e.g][]{haywood06}, though 
\cite{holmberg07} claim that GCS $uvby$ photometry of stars belonging to the 
Hyades cluster is not on the same scale as the rest of the catalogue, possibly 
because they were observed at higher air masses from Chile. In the case of 
Hyades, its controversial $c_1$ colour anomaly is also of concern 
\citep[i.e.~ the systematic difference in the $c_1$ vs.~$(b-y)$ diagram 
between the sequence of unevolved stars in Hyades and the corresponding 
sequence for unevolved Coma and field stars with similar $\delta m_1$; see 
e.g.,][for more details]{c75,stromgren82}. 

We took Str\"omgren photometry for the Hyades cluster from
\cite{cp66}\footnote{The extension of the original $uvby$ system to
  cool and metal-poor stars is based on two main sets of standard
  stars, those of \cite{bond80} and \cite{olsen93}, respectively. The
  main discrepancy between the two concerns the $c_1$ index, stemming
  from differences in $u$ band \citep[see discussion
  in][]{olsen95}. Our metallicity calibration uses GCS photometry,
  which is built on the Olsen standards \citep[see][and references
  therein]{nordstrom04}. For testing in the metal-rich regime, as we
  do here, these differences are of no concern since the original set
  of observations defining the $uvby$ system (and therefore adopted
  also by Olsen) is used.} and compared the result of our metallicity
calibration with the detailed spectroscopic study of Hyades
stars of \cite{paulson} (who adopt a $\teff$ scale rather close to
our IRFM scale). There are 10 single stars in common and we find a
mean $\feh = 0.09 \pm 0.02$~dex ($\sigma=0.06$~dex). This value is
slightly lower than the mean obtained using the same stars in
\cite{paulson}, which amounts to $\feh=0.14 \pm 0.01$~dex
($\sigma=0.04$~dex) and nearly coincides with the mean value derived
by \cite{paulson} using a larger sample of cluster members. We note
that for this cluster a typical metallicity around $0.1$~dex is
commonly cited in the literature \citep[e.g.,][]{tj04,schuler06}.

The Hyades open cluster is known to be underabundant in helium for its 
metallicity 
\citep[$\Delta Y \sim 0.02$ see e.g.,][and references therein]{vandenberg10}, a 
feature which would be tempting to associate to the $c_1$ colour anomaly 
\citep{stromgren82}. However, synthetic colours show that variations of helium 
of this order affect $c_1$ to a negligible extent \citep{melendez10}. 
Another possibility is that the anomaly is caused by variations in other 
elements.
In fact the Hyades anomaly could simply be the $\feh$ difference between stars 
of similar $\delta m_1$ as the following comparison with the Coma cluster 
suggests. For 17 stars in the Coma cluster we took the photometry of 
\cite{cb69} and derived $\feh=-0.08 \pm 0.02$ ($\sigma=0.07$~dex) using our 
calibration, which implies a metallicity difference with respect to Hyades that 
excellently agrees with that spectroscopically measured by \cite{bf90} and 
\cite{fb92}. 

Another open cluster originally observed by \cite{cb70} is NGC752. This 
cluster also has relatively low reddening $E(b-y)=0.027$ 
\citep{att06}. Using all dwarfs in \cite{cb70} (within the colour range of our 
calibration) gives $\feh=-0.07 \pm 0.02$~dex ($\sigma=0.11$~dex), and a similar 
value ($-0.05 \pm 0.04$~dex, $\sigma=0.09$~dex) when restricting the same 
photometric measurements to the smaller --yet with cleaner membership-- sample 
of \cite{att06}. 

Finally, using observations of F-type stars in the Pleiades (which are less 
affected by activity stemming from the young age of this open cluster) from 
\cite{cp76} and adopting $E(B-V)=0.04$ \citep[e.g.,][]{vanLeeuwen09}, we derive 
$\feh=0.00 \pm 0.02$~dex ($\sigma=0.10$). The difference with respect to the 
Hyades again excellently agrees with that obtained from the 
spectroscopic comparison of \cite{bf90}, after correcting the Pleiades for 
known non-members \citep{an07}. Our 
$\feh$ also agrees well with recent spectroscopic estimates based on a 
$\teff$ scale consistent with our own \citep{soderblom09}. For the last two 
clusters, we also checked that a typical uncertainty of $E(b-y)=0.01$ 
affects $\feh$ by $\sim 0.01$~dex. 

\begin{figure}
\includegraphics[scale=0.74]{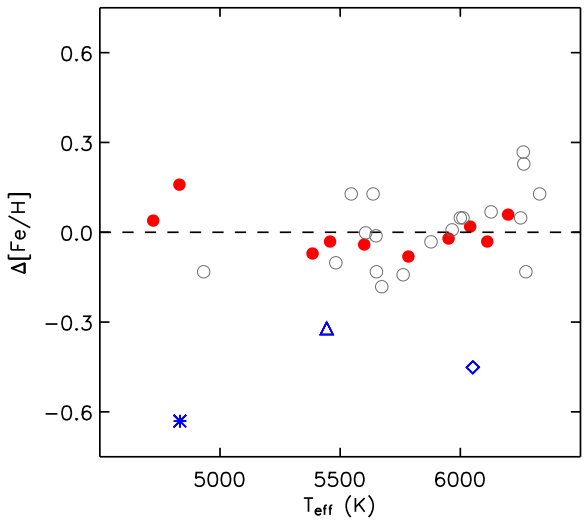}
\caption{Differential Fe abundances of kinematically selected members of 
HR1614. Filled circles are metal-rich members, while asterisk (HIP13513), 
triangle (HIP6762) and diamond (HIP25840) are spurious members according to the 
chemical tagging performed by \cite{desilva}. Open circles are the analogous 
comparison with the group members from \cite{fh00}.}
\label{f:hr1614}
\end{figure}

An additional check on the precision of our metallicity calibrations comes 
from the HR1614 moving group \citep{eggen78,fh00}. Chemical tagging via 
high-resolution 
spectroscopy of kinematically selected members allows us to clearly identify 
interlopers amongst the group members \citep{desilva}. 
Fig.~\ref{f:hr1614} shows the differential Fe abundance $\Delta \feh$ for a 
number of candidate members
in common between \cite{desilva} and GCS, using our metallicity calibration. 
The plot is relative to the mean metallicity of 
the sample, and thus largely independent on the underlying $\teff$ scale 
adopted. The comparison agrees remarkably well with figure 2 in 
\cite{desilva}, clearly allowing us to identify spurious members of the moving 
group. We determine the group to have a mean $\feh=0.28 \pm 0.02$~dex 
($\sigma=0.07$~dex), in good agreement with the spectroscopic value of 
$0.25$~dex in \cite{desilva}. 

\begin{figure*}
\includegraphics[scale=1.2]{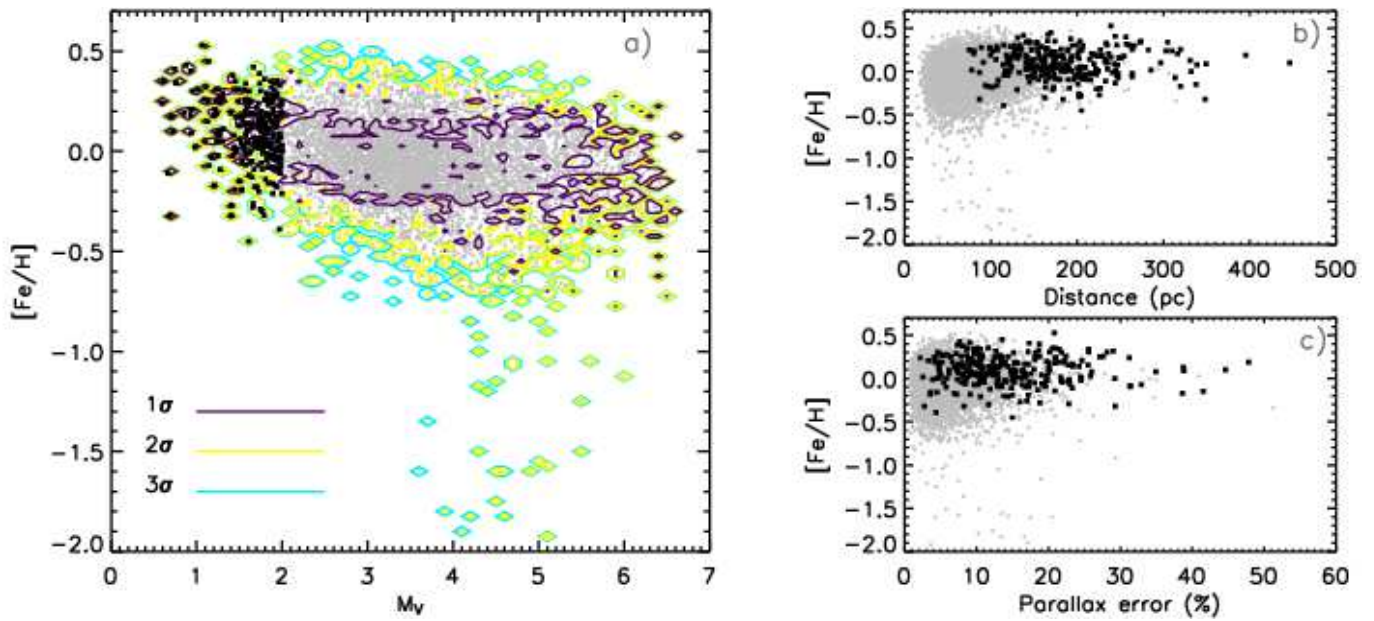}
\caption{Panel a): $\feh$ distribution when slicing in absolute magnitude 
($M_{V_T}$) 5976 stars (grey circles) that belong to the {\it irfm} sample and 
are within the metallicity calibration range. Contour levels are computed on 
abscissa intervals of $0.1$~mag to equally represent regions with fewer stars. 
Panel b) and c): $\feh$ distribution of the 
same stars, but plotted as a function of distance and parallax error. In all 
panels, stars with $M_{V_T}<2$ are overplotted as black squares (271 in total).} 
\label{f:trends}
\end{figure*}

Finally, we comment on the accuracy of our photometric metallicities
for intrinsically bright stars. The spectroscopic sample upon
which our calibration is built extends to magnitudes only slightly
brighter than $M_{V_T}\sim 2$, which are typical for F dwarfs; however,
the GCS contains some hundreds of stars more luminous than this (also
compare with Fig.~\ref{fig:HRteff}). These stars are close to the
instability strip and are therefore possibly contaminated by $\delta$
Scuti pulsators and/or chemically peculiar A/F stars \citep[the latter
often being overabundant in Fe and possibly with peculiar colours,
e.g.,][]{gmr08,neto08}. Our calibrations include a dependence on the
$c_1$ index (a good surface gravity indicator for hot stars), so in
principle, we can expect them to work for decreasing $\logg$.
Fig.~\ref{f:trends} shows a clear trend for the brightest stars in the
GCS, which tend to be more metal-rich than the remaining part of the
sample (left panel). Because they are preferentially metal-rich, an age
determination based on isochrones also biases them to even younger
values. These bright stars also stand out in the study of the
metallicity distribution function (Section \ref{sec:mdf}) and in kinematic 
(Fig.~\ref{fig:agevelsig}).  Some uptrend in this figure
(starting around $M_{V_T}\sim 3$) is expected from
colour/spectral-type cuts in the original GCS sample selection, because at a 
given colour metal-poor dwarfs are fainter (cf.~also Fig.~\ref{f:agemass}). 
However, at the bright end (in particular from $M_{V_T}\sim 2$) the calibration 
seems to deviate too strongly. Interestingly, those stars are preferentially 
the most distant ones and thus have increasing reddening uncertainties as
  well as the largest parallax errors, which could misleadingly place
  intrinsically bright stars at fainter absolute magnitudes. Also
  notice that by sampling larger distances, where the GCS is not complete
  anymore, intrinsically luminous stars are preferentially found
  around the peak of the metallicity distribution function, which
  could partly account for their rather high $\feh$.  In addition,
  preferentially higher metallicities could also stem from these
  objects originating from the inner disc (cf.~Section \ref{sec:disc}),
  a feature of which we do not find any significant indication,
  though. Even if not conclusive, this seems to suggest that
removing stars with an absolute magnitude $M_{V_T}$ brighter than $2$ would
be a safe choice when using GCS stars for deriving local constraint on
Galactic chemical evolution.

\subsection{The mild sensitivity of Str\"omgren photometry to the $\alpha$-elements}\label{sec:alpha}

For all 1498 calibrating stars presented in Section \ref{sec:feh} we have 
$\feh$ and $\aFe$ from high-resolution spectroscopy. From this the overall 
metal--to--hydrogen ratio $\mh$ can be computed \citep[e.g.,][]{yi01}. 
Interestingly, when fitting functional forms of the kind of 
Eq.~(\ref{eq:m1_feh}) and (\ref{eq:c1_feh}) to $\mh$, the scatter of the 
resulting calibrations decreases to $0.07$ and $0.10$~dex respectively, thus 
suggesting \citep[cf.~e.g.,][]{yong08} that Str\"omgren indices carry 
information on the overall metal content (apart from a few of the most 
metal-poor stars in the sample, which indicates diminishing sensitivity 
to $\mh$ because of the intrinsically fewer lines). 
\begin{figure*}
\includegraphics[scale=0.72]{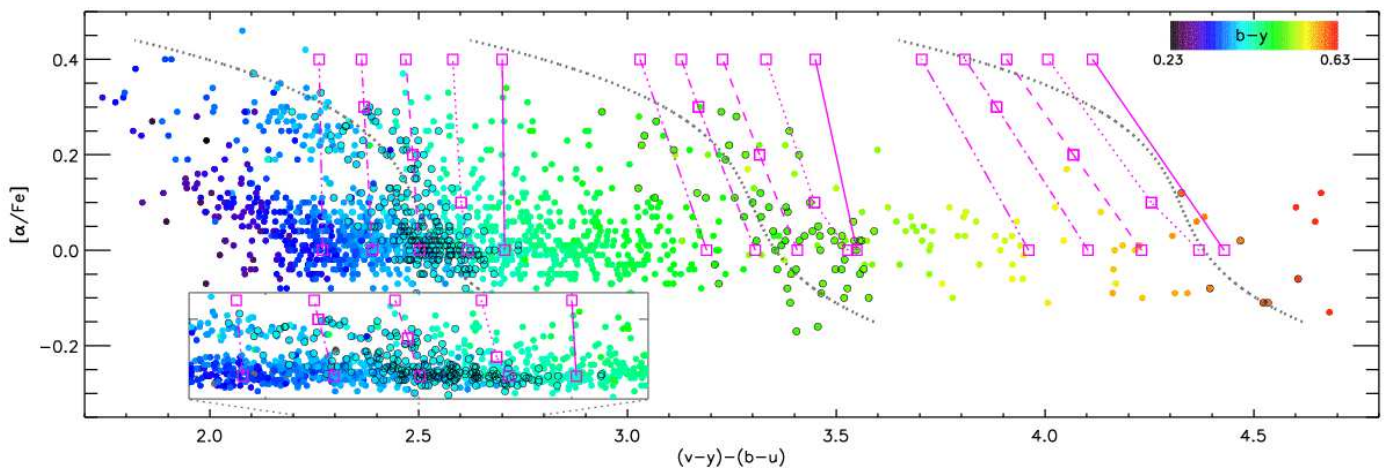}
\caption{$\aFe$ versus $(v-y)-(b-u)$ for our 1498 calibrating 
stars. Bluer (redder) colours indicate hotter (cooler) stars, 
according to their $(b-y)$, as shown in the top right box. Squares 
are synthetic colours computed from MARCS model fluxes at fixed 
$\aFe$ (as available from MARCS library) for selected values of 
$(b-y)=0.4,\,0.5,\,0.6$ (from left to right) and 
$\feh=0.00$~dex with $\aFe=0:0.4$~dex (continuous line)
$\feh=-0.25$~dex with $\aFe=0:0.1:0.4$~dex (dotted lines), 
$\feh=-0.50$~dex with $\aFe=0:0.2:0.4$~dex (dashed lines), 
$\feh=-0.75$~dex with $\aFe=0:0.3:0.4$~dex (dot-dashed lines), 
$\feh=-1.00$~dex with $\aFe=0:0.4$~dex (triple-dot-dashed lines). 
Stars within 0.01 mag of the selected $(b-y)$ interval are shown 
with open circles to highlight the trend. Grey dotted lines are 
fiducials built for those stars. Lower left panel is a zoom of the 
$(b-y) = 0.4$ data set for $2.2 \le (v-y)-(b-u) \le 2.8$.}
\label{f:alpha}
\end{figure*}

It would be possible to apply our calibrations as a function of $\feh$ and 
$\mh$ and from those derive an estimate of $\aFe$. In practice 
though, there is some degree of correlation in the results since the same 
functional form and indices are used over 2~dex in metallicity to estimate 
typical alpha-enhancements within $\sim 0.5$~dex. 
We experimented with different combinations of Str\"omgren colours and found 
$a_1=(v-y)-(b-u)$ to be sensitive\footnote{Also other indices have been found 
to show some dependence on $\aFe$ such as e.g.,~$m_1-(b-y)$. From our 
investigation it seems that Str\"omgren filters such as $b$ and $y$ are barely 
affected by $\aFe$, while $u$ and $v$ are more affected.} 
to $\aFe$ at a given $\teff$. 
Fig.~\ref{f:alpha} shows $\aFe$ versus our index $a_1$ as well as the 
comparison with synthetic colours at a few $(b-y)$ values for the sets of 
alpha-enhanced and -poor models available through the MARCS library.

In Fig.~\ref{f:alpha} a dependence on $\feh$ is certainly built in given that 
stars with lower $\feh$ have preferentially higher levels of 
alpha-enhancements. 
Nevertheless, the comparison with synthetic colours shows that the trend is 
real at fixed metallicities. As we already pointed out, Str\"omgren 
synthetic colours are not immune to deficiencies, and combinations 
of the adopted filters are --by construction-- also sensitive to 
metallicity and surface gravity. 
Changing the latter parameter shifts synthetic colours to the right 
or left with respect to the position shown in Fig.~\ref{f:alpha}, 
which refers to $\logg=4.5$. However, we checked that the shape of 
the slopes remains unaffected by the exact value of $\logg$. 
Limitations in synthetic colours as well as surface gravity 
dependence could explain why the bulk of calibrating stars is fitted by 
models having $\feh=-0.5$ rather than a higher metallicity, which is more 
representative of the sample (cf.~also with Fig.~\ref{f:m1}). 
According to the models, in Fig.~\ref{f:alpha} the sensitivity to 
$\aFe$ is more pronounced (i.e.~it has a shallower slope) at cooler 
effective temperatures, which are therefore likely to be better 
recovered. Determining $\aFe$ becomes increasingly difficult for 
the hottest and most metal-poor stars, as expected because both 
atomic and molecular lines get weaker in this regime 
\citep[e.g.][]{coelho05}. Yet, even at the bluest colours the data 
seem to show a clearer trend with $\aFe$ than models. Aware 
of these warnings, the mild correlation of the $a_1$ 
index with alpha-enhancement seems to work for drawing meaningful 
conclusions when one has a statistically large sample of stars 
(see also Fig~\ref{f:rl}).

\begin{figure}
\includegraphics[scale=0.95]{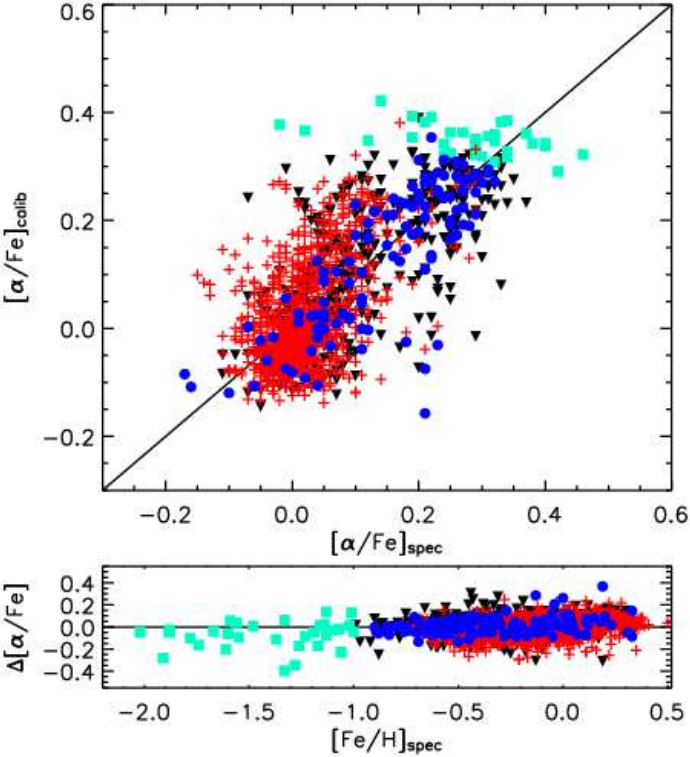}
\caption{Upper panel: spectroscopic versus Str\"omgren $\aFe$ estimates for 
our 1498 calibrating 
stars. Crosses (circles) are stars having probability $>90$ percent of being 
thin (thick) disc based on their kinematic (using $U,V,W$ velocities from the 
GCS). Downward triangles are stars with lower probability or 
for which kinematic information was not available. Filled squares are stars 
having $\feh \le -1$ (independently of their kinematic thin/thick membership,  
if available). Lower panel: same symbols as above, showing the difference 
spectroscopic minus ours.}
\label{f:alpha2}
\end{figure}

\begin{figure}
\includegraphics[scale=0.64]{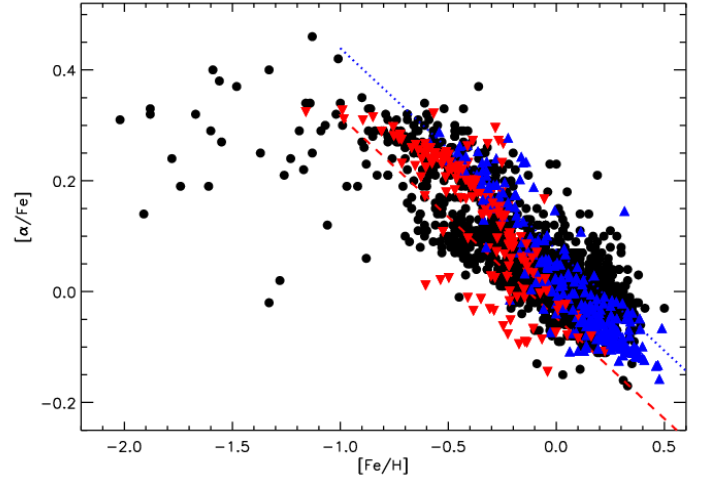}
\caption{$\feh$ versus $\aFe$ for the 1498 stars in the spectroscopic 
sample (filled circles). Dotted and dashed lines are arbitrarily used to 
select stars on the right and left locus of the plot (only for $\feh > -1$). 
Downward (upward) triangles represent stars selected according to this 
criterion, but using our $\feh$ and $\aFe$ Str\"omgren calibrations.}
\label{f:rl}
\end{figure}

For each $(b-y)$ in Fig.~\ref{f:alpha} we constructed a fiducial using 
stars of similar $\teff$ and derived a value of $\aFe$ according to 
their $(v-y)-(b-u)$ with respect to that of the corresponding 
fiducial. Despite models show a spread with metallicity in Fig.~\ref{f:alpha}, 
we did not include any dependence on $\feh$ in building 
the fiducial to avoid any risk of introducing a spurious trend of 
increasing alpha with decreasing $\feh$. The comparison 
between the spectroscopic measurements and our photometric estimates is shown 
in Fig.~\ref{f:alpha2}. The overall agreement is indeed good (formally 
$\sigma=0.09$~dex), though there are a few caveats: $\aFe$ for stars with 
$\feh \lesssim -1$ is not well recovered (the calibration saturates, filled 
squares), as expected from our previous discussion on metal-poor stars. Also, 
for thin-disc stars $\aFe$ tends to be slightly underestimated/overestimated 
at higher/lower metallicities. This reflects the shape of the fiducials used 
to derive $\aFe$ from Fig.~\ref{f:alpha}. In fact, 
Fig.~\ref{f:rl} shows that our $\aFe$ 
calibration does not allow us to recover any gap between thin and thick 
disc stars. The shape of the overall narrow trend is thus driven from 
the fiducial, yet within this trend a distinction between alpha-rich and 
-poor stars is possible, though in a statistical sense only. This is shown by 
selecting calibration stars on the right (left) of the dotted (dashed) line in 
Fig.~\ref{f:rl}, with the same distinction right vs.~left still being preserved 
when we use our $\feh$ and $\aFe$ calibrations, which are represented by upward 
vs.~downward triangles. 
Notice that using our $\aFe$ the dispersion of the photometric $\mh$ 
with respect to the spectroscopic measurements is $0.08$~dex, compared to 
$0.10$~dex (previous section) when using $\feh$ only.

In conclusion, the sensitivity of our approach to the alpha elements is real, 
but mild and works only for $\feh > -1$ or slightly higher values. 
Also, a statistical distinction between alpha-rich and -poor stars is 
possible, but only within the functional form of our calibration so 
that other finer structures could still be missing. Thus, the values 
of alpha-enhancements we derive are not exact measurements of $\aFe$, but 
rather 
a proxy of them for stars of similar $\feh$. For this reason we will refer to 
our estimate as $\alfe$ instead of $\aFe$ throughout the paper. 
Nevertheless, as we show below, when one deals with several thousands of 
stars, as is the case in the GCS, our $\alfe$ can give important insight into
the formation and evolution of the Galactic disc(s).

\section{New age and mass determinations}\label{sec:age}

\begin{figure*}
\epsfig{file=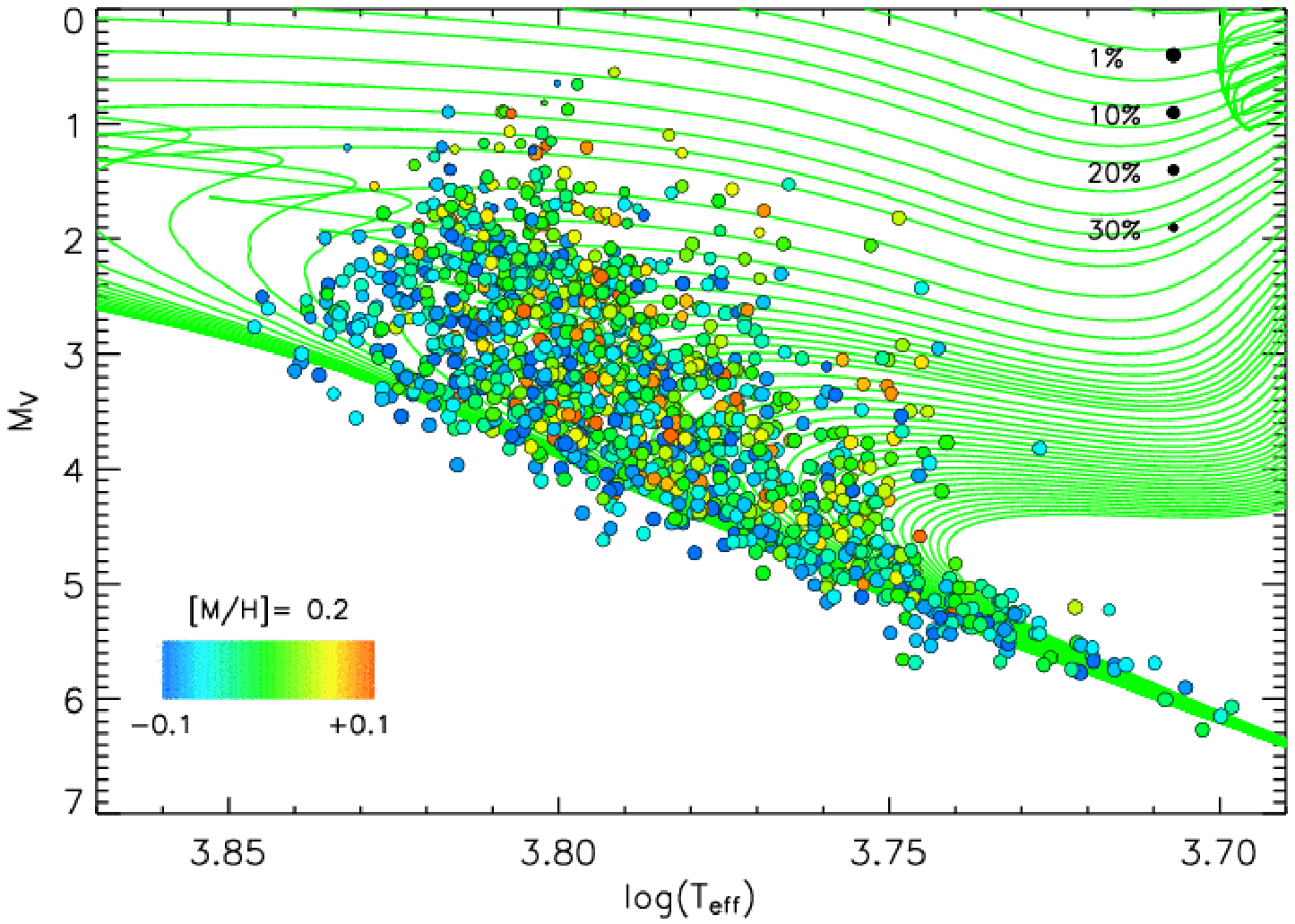 ,width=0.5\hsize}
\epsfig{file=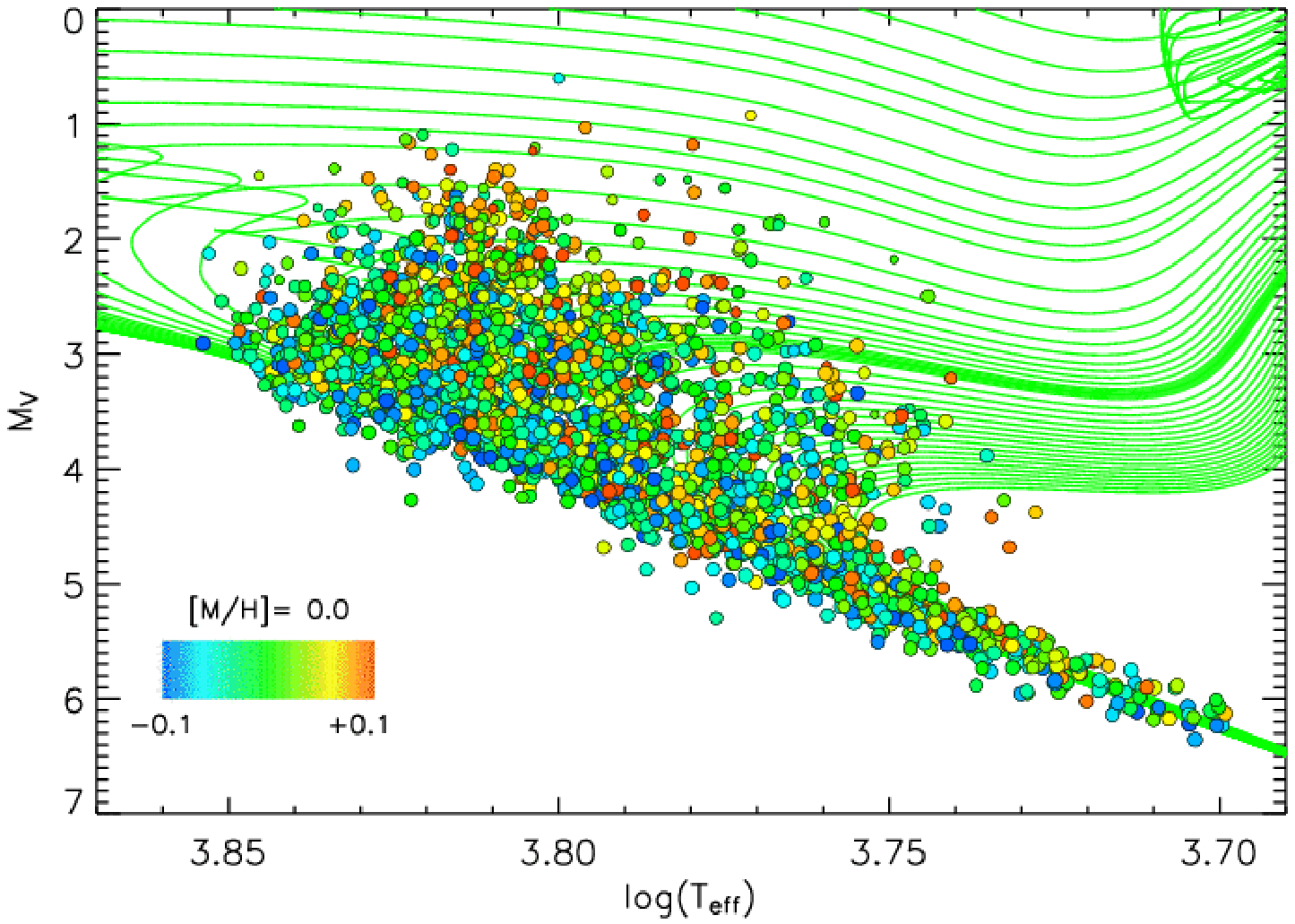 ,width=0.5\hsize}
\epsfig{file=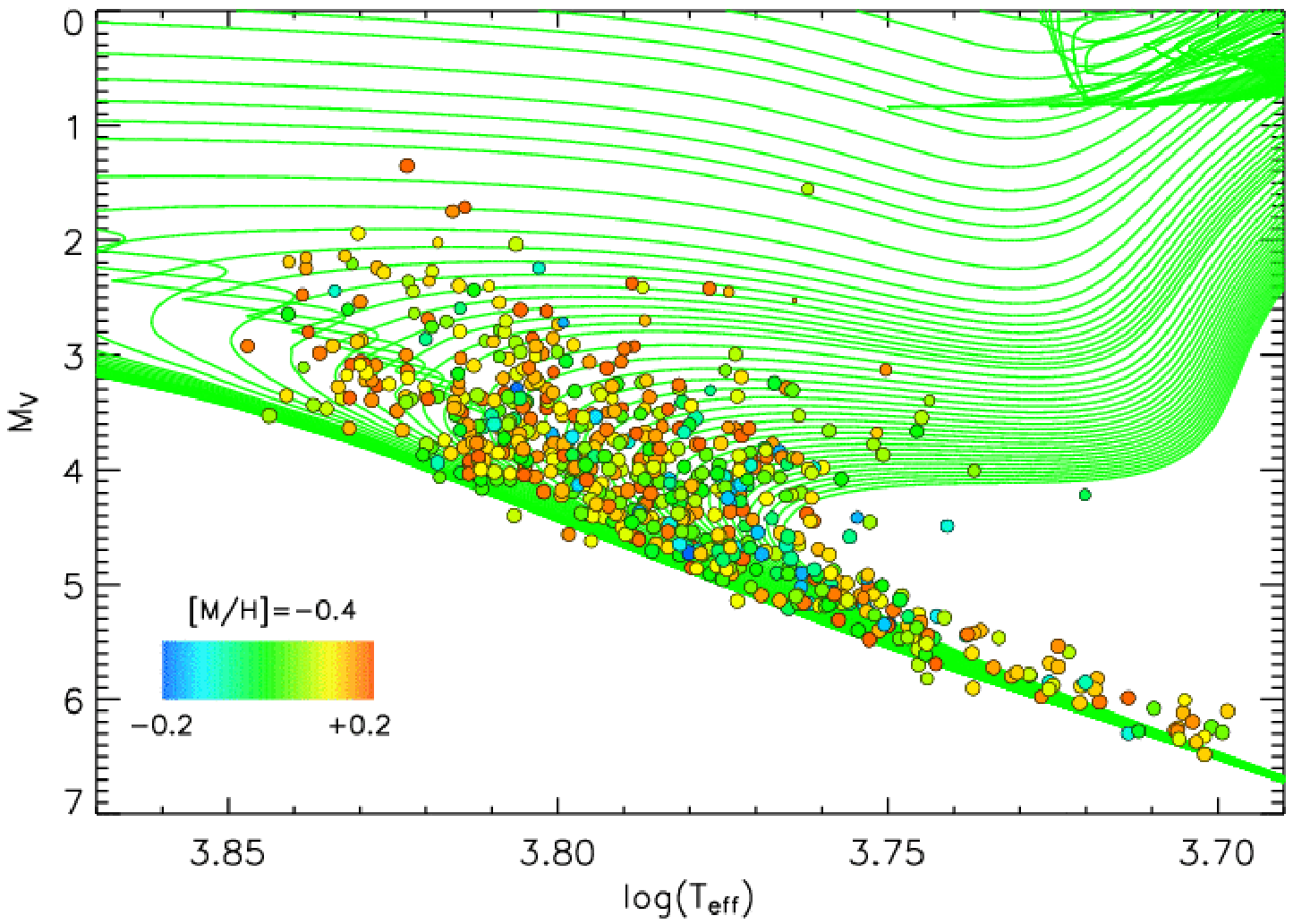 ,width=0.5\hsize}
\epsfig{file=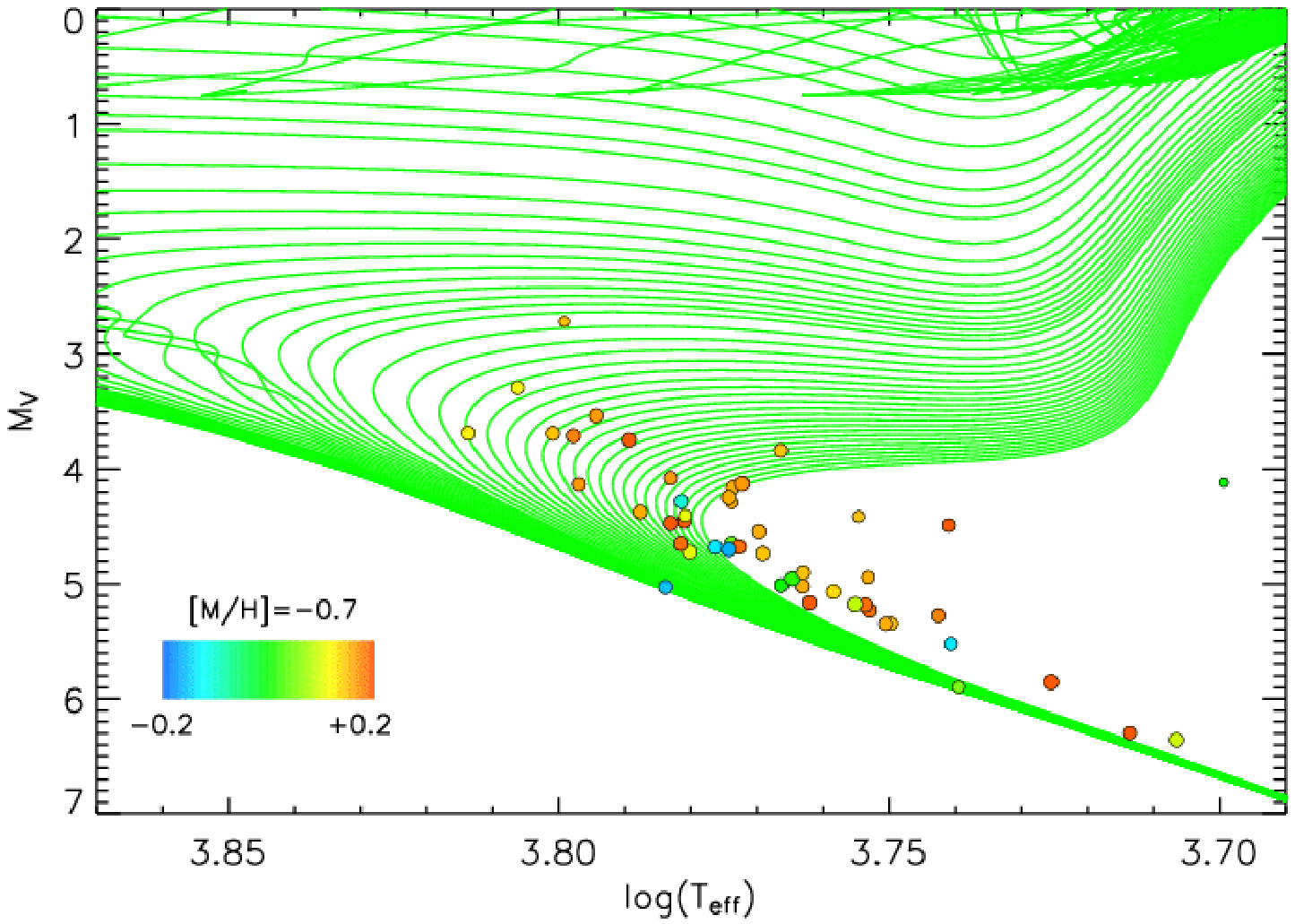 ,width=0.5\hsize}
\caption{BASTI isochrones for different ages at a given metallicity 
(continuous lines) compared to stars of similar $\mh$, where the difference 
$\pm 0.1$ or $\pm 0.2$~dex is coded by colours. Larger symbols are for stars 
with higher parallax accuracy as labelled in the top left panel. Only stars in 
the {\it irfm} sample are shown.}\label{fig:HRteff}
\end{figure*}

Revising metallicities and effective temperatures also affects age and
mass estimates for the stars.  \figref{fig:HRteff} shows comparisons
between isochrones and stars with metallicities close to those of the
plotted isochrones. Compared to previous studies, our improved
effective temperatures are hotter, and the large systematic
discrepancies between theoretical isochrones and observed data that
plagued e.g.,~\cite{Pont04} almost entirely disappear.  As can be
seen, only at the lowest metallicities and luminosities the
theoretical main sequence has the tendency to fall beneath the stars,
i.e.,~the isochrones are too hot. However, the discrepancy is
considerably reduced from earlier GCS analyses where shifts in the
effective temperature of the isochrones have to be introduced below
solar \citep{nordstrom04} or even at all \citep{holmberg07}
metallicities\footnote{Note that in GCSII also the solar isochrone,
  which is {\it constructed} to fit the Sun, has to be cooled by
  $0.005$~dex in $\log \teff$, corresponding to approximately $70$~K,
  in agreement with the offset in Fig.~\ref{f:teff}.}.  Our new
temperature and metallicity scales prove to agree very well with those of 
theoretical isochrones, at least for metallicities
higher than about $\mh>-0.5$, which includes the vast majority of
stars in our sample (cf.~Fig.~\ref{f:mdf}). Indeed, for the sake of
Fig.~\ref{fig:HRteff} a metallicity bias stemming from the wings of
the metallicity distribution function will also play a role, as we
discuss in greater detail in the appendix, as well as the monotonic decrease 
(increase)
of stars with metallicity in the metal-rich (-poor) tail of the
metallicity distribution function by which the average metallicity in
a given interval can be lower (higher) than the middle value of the
interval. This can be clearly seen in the net bias of hotter and more
metal-poor (cooler and more metal-rich) stars in the top left (bottom
right) panel of Fig.~\ref{fig:HRteff}, where stars are selected in
symmetric intervals around the metallicity of the
isochrones\footnote{At the same time, a stronger disagreement for
  metal-poor low main sequence stars --for which their position on the
  HR diagram is substantially age independent-- was noticed by
  \cite{casagrande07} and it could have potential implications for
  studies of multiple stellar populations \citep{portinari10}. The
  same $\teff$ scale adopted here compares well with isochrones for
  nearby, evolved subdwarfs, suggesting that this disagreement seems now
  reduced at least for $\log \teff >3.7$ \citep[compare with fig.~10
  in][]{vandenberg10}, though further investigations are encouraged.}.

As pointed out by \cite{Pont04}, na\"ive fits to isochrones lead to
severe biases, e.g.~what they name a terminal age bias. This happens
because some places on isochrones are more densely populated than
others because of the mapping from mass to colours/luminosity owing to the
initial mass function and to the time scales involved in stellar
evolution. Just looking for the closest match ignores these facts and
might erroneously place too many stars into sparsely populated
regions. Biases of this kind can be accounted for by taking a Bayesian
approach as in \cite{Pont04} and \cite{Jorgensen05}, who did a Bayesian
age determination on the old GCSI-II. A detailed discussion can also
be found in \cite{bb10}. In our sample the errors vary significantly
between stars, but they depend only weakly on the derived stellar
parameters, so that we can neglect this influence on the age
distribution. We only used $\log(\teff)$, absolute Johnson $V$ magnitude and
metallicity information to estimate the ages and masses of stars. In
principle more information could be in the colours, but essentially
this is already exploited by the colour-dependent
calibrations. Moreover, a direct use of colour information would imply
relying directly on synthetic colours \citep[with the $uvby$ bands being
more troublesome than others, see
e.g.,][]{onehag09,arnadottir10,melendez10}, which we wanted to
avoid. Further, especially Str\"omgren colours depend on $\mh$, thus
requiring an even denser grid of metallicities than the one we use.

The isochrones provide us with a natural grid for calculating the probability distribution function for the parameters of a given star. Every isochrone point has to be weighed by the volume of parameter space it has to cover and by the a priori assumptions. To avoid any factor that could contribute to the particular age distribution that we find in the sample, we assume a flat age prior for $0-14$~Gyr, i.e. a constant density ${\mathcal A}({\tau})$ of 
stars over age. 
\begin{displaymath}
{\mathcal A}({\tau}) = \left\{ 
\begin{array}{lll} 
1 & for & 0 \leq \tau \leq 14\, \Gyr \\ && \\ 0 & else
\end{array}
\right.
\end{displaymath}
As different positions of stars in the Hertzsprung-Russell diagram
imply different underlying selection functions that bias the intrinsic
age distribution at this place, this approach also avoids making
further assumptions that could potentially weaken the interpretation
of the results. The mass prior is a Salpeter IMF \citep{salpeter55}
and we do not set any dependence on age. In the mass interval of
  interest here (cf.~Fig.~\ref{f:agemass}), a Salpeter IMF is indeed 
  still appropriate, whereas considerably larger uncertainties exist
  regarding the lower and higher mass range
  \citep[e.g.,][]{bastianARAA}.  We further tested the extreme case of
  a flat IMF, and even this unrealistic assumption has negligible
  impact on our results (see the appendix, where we provide details on
the Bayesian scheme adopted for dealing with the observational errors
in $\teff$, metallicities and absolute magnitudes).

In order to study differences between different isochrones, we used grids of the BASTI\footnote{http://www.oa-teramo.inaf.it/BASTI} \citep[][]{pietrinferni04,pietrinferni06,pietrinferni09} and Padova\footnote{http://stev.oapd.inaf.it/YZVAR/cgi-bin/form} isochrones \citep[][]{bertelli08, bertelli09}. The Padova grid has a logarithmic age spacing of $0.01$~dex, i.e. it rises from $23$ Myr at $\tau = 1$~Gyr to $230$~Myr at $\tau = 10$~Gyr. We queried a total of 56 metallicities from the database, which are created by interpolating among the nine available metallicities, ranging from $Z=0.0001$ to $Z=0.07$. The solar isochrone has $Y_{\odot} = 0.26$ and $Z_{\odot} = 0.017$ \citep{grevesse98}. The helium-to-metallicity enrichment ratio was chosen to be $\Delta Y / \Delta Z = 2.1$, which is consistent with the value inferred from the study of metal-rich local K dwarfs \citep{casagrande07}. At the lowest metallicities this falls somewhat short of $Y = 0.23$ \citep[the lower helium abundance in the database, lower than the current preferred estimate from $\rm{WMAP}+\rm{BBN}$, see e.g.,][]{steig10}, and for those objects we kept this value of $Y$. 
For the BASTI isochrones \citep[$Y_{\odot} = 0.2734$ and $Z_{\odot} = 0.0198$ from][]{grevesse93} we used a denser grid than the published one. This grid was specially calculated for this purpose to include 20 metallicities at $\Delta Y / \Delta Z = 1.45$ (leading to a primordial helium abundance in agreement with the cosmological estimate) in the range $Z=0.0001$ to $0.04$, with a time spacing of $100$~Myr maximum, making this grid denser (sparser) at high (lower) ages compared to the logarithmic age spacing in the Padova isochrones. Both sets of isochrones assume solar-scaled abundances (i.e.~constant ratio of the single metals with respect to the Sun), which are appropriate because it has been shown that for the range of metallicities covered by the present study, isochrones for enhanced $\alpha$ abundances can be reproduced remarkably well by those for solar scaled mixtures if $Z$ is the same \citep[e.g.,][]{chieffi91,csd92,sw98,v00}. We 
also checked the difference when using $\feh$ rather than $\mh$ 
in determining ages: the overall difference is fairly small, with a 
scatter of about $0.5$~Gyr.

\begin{figure}
\includegraphics[scale=0.52]{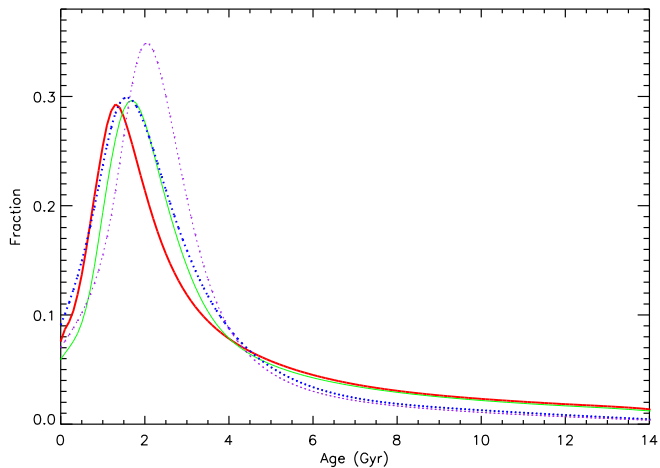}
\caption{Normalized age probability distribution for all stars in the sample having both BASTI (continuous thick) and Padova (continuous thin) ages and only for stars in the {\it irfm} sample (dotted thick and thin lines) having good ages and $M_{V_T} \ge 2$.} 
\label{fig:agedist}
\end{figure}

\figref{fig:agedist} shows the age probability distributions for all stars in the GCS with ages determined from both BASTI and Padova isochrones and also for all stars with good ages. Throughout the paper, ages are defined to be good if $\sigma<1$~Gyr or the relative uncertainty is better than 25 percent (see also the appendix). While these criteria are arbitrary, they balance a reasonable determination of absolute ages for young objects with a reasonable relative determination for older ones. The distribution strongly peaks around 2~Gyr, which is caused by the selection effects on the sample \citep[see also][]{nordstrom04}. 

The GCS is in fact limited near the plane of the disc, while older stars usually have a considerably more extended vertical distribution, which brings their orbits high above the plane and thereby lowers their presence in this survey. In addition, the magnitude limits of the catalogue give a larger volume to bright, young stars, and the exclusion of giant and very blue stars from the sample leads again to a net bias against very young and especially against old objects. An estimate of the age of the disc can thus not be done directly using the age of the stars in the present sample, but requires modelling the star-formation history of the solar neighbourhood, returning a considerably older disc 
\citep[$>10.5$~Gyr, see e.g.,][]{ab09,ralph09a}.
Also notice that because of the young ages, our sample is fairly immune to 
atomic diffusion, possibly apart for a few of the oldest stars. 

  Fig.~\ref{f:agemass} clearly summarizes all main issues in dating
  stars. Ages are most readily determined for stars in the upper
  envelope of Fig.~\ref{f:agemass}, which roughly maps the turn-off
  region. At low masses, apart from the most metal-poor subdwarfs,
  reliable ages are difficult to derive (grey dots) because the
  majority of these stars are still on the main sequence due to their
  long lifetimes. In addition, somewhat below $1\,M_{\odot}$ the GCS
  starts losing completeness, being mostly limited to FG dwarfs. The
  youngest stars cover a short mass range (cf.~blue points in the
  middle panel of Fig.~\ref{f:mdfAGE}): more massive young stars are
  in fact brighter and hotter than sample selection limits, apart from
  a handful of bright objects (squares). The reliability of our
  metallicity calibration for those stars was already discussed
  in Section \ref{sec:further}. The depletion of stars longward of the
  kink at $\sim 1.5\,M_{\odot}$ precisely suggests that at masses
  higher than this value the sample is partly incomplete \citep[which
  roughly corresponds to $M_{V_T} \sim 2$, using the mass-luminosity
  relation of e.g.][]{henry93,ff10}. On the contrary, no obvious
  biases seems to be present in the range $1.1 \lesssim
  \frac{M}{M_{\odot}} \lesssim 1.5$.
\begin{figure}
\epsfig{file=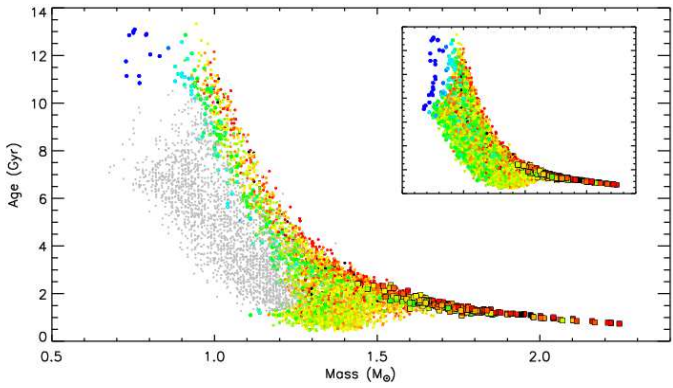 ,width=1\hsize}
\caption{Ages versus masses for stars belonging to the {\it irfm} sample. 
Colours are for stars with well determined ages, going from metal-poor (blue) 
to -rich (red), while grey dots are for the remaining stars. Squares are stars 
brighter than $M_{V_T}=2$. Inner panel: same as outer panel, but with a metallicity 
colour coding also for stars with less reliable ages.} 
\label{f:agemass}
\end{figure}

\section{The metallicity distribution function}\label{sec:mdf}

Given its complete nature (see Section \ref{sec:intro}), the GCS is well 
suited for the study of the metallicity distribution function (MDF) in the 
solar neighbourhood \citep{nordstrom04,holmberg07}. However, this does not 
mean that the MDF shown here can be directly compared to theoretical 
expectations. For the same reasons presented above when 
discussing the age distribution, sample selection effects enter the results. 
For a quantitative comparison those selection effects have to be taken into 
account in theoretical models \citep[cf.][]{ralph09a}.
\begin{figure}
\includegraphics[scale=0.35]{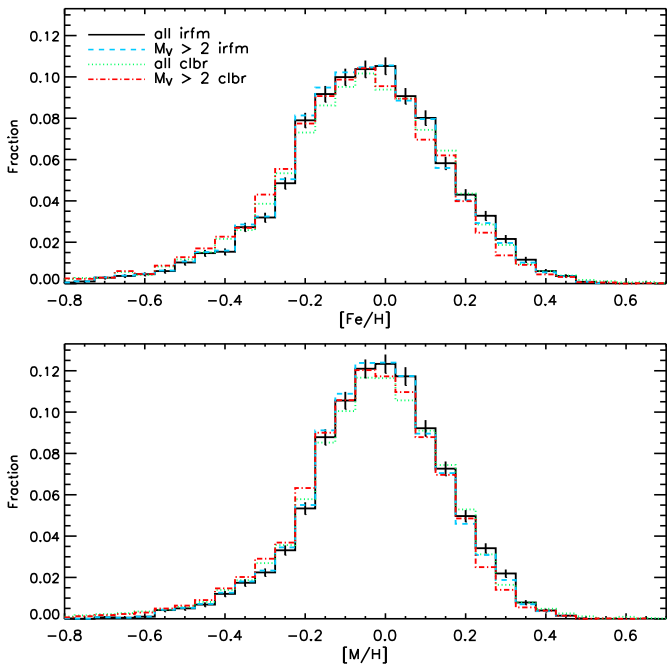}
\caption{MDF of the solar neighbourhood in terms of $\feh$ (upper panel) and  
$\mh$ (lower panel). 
Continuous line refers to stars belonging to the {\it irfm} sample (5976 stars 
within the colour ranges of the metallicity calibration), dashed line when 
considering only stars fainter than $M_{V_T}=2$, dotted line to all {\it clbr} 
stars (8470 within the colour ranges of the metallicity calibration) and 
dot-dashed when applying the same luminosity cut as above. Poisson error bars 
are shown for a representative case in both panels.} 
\label{f:mdf}
\end{figure}

\begin{figure}
\epsfig{file=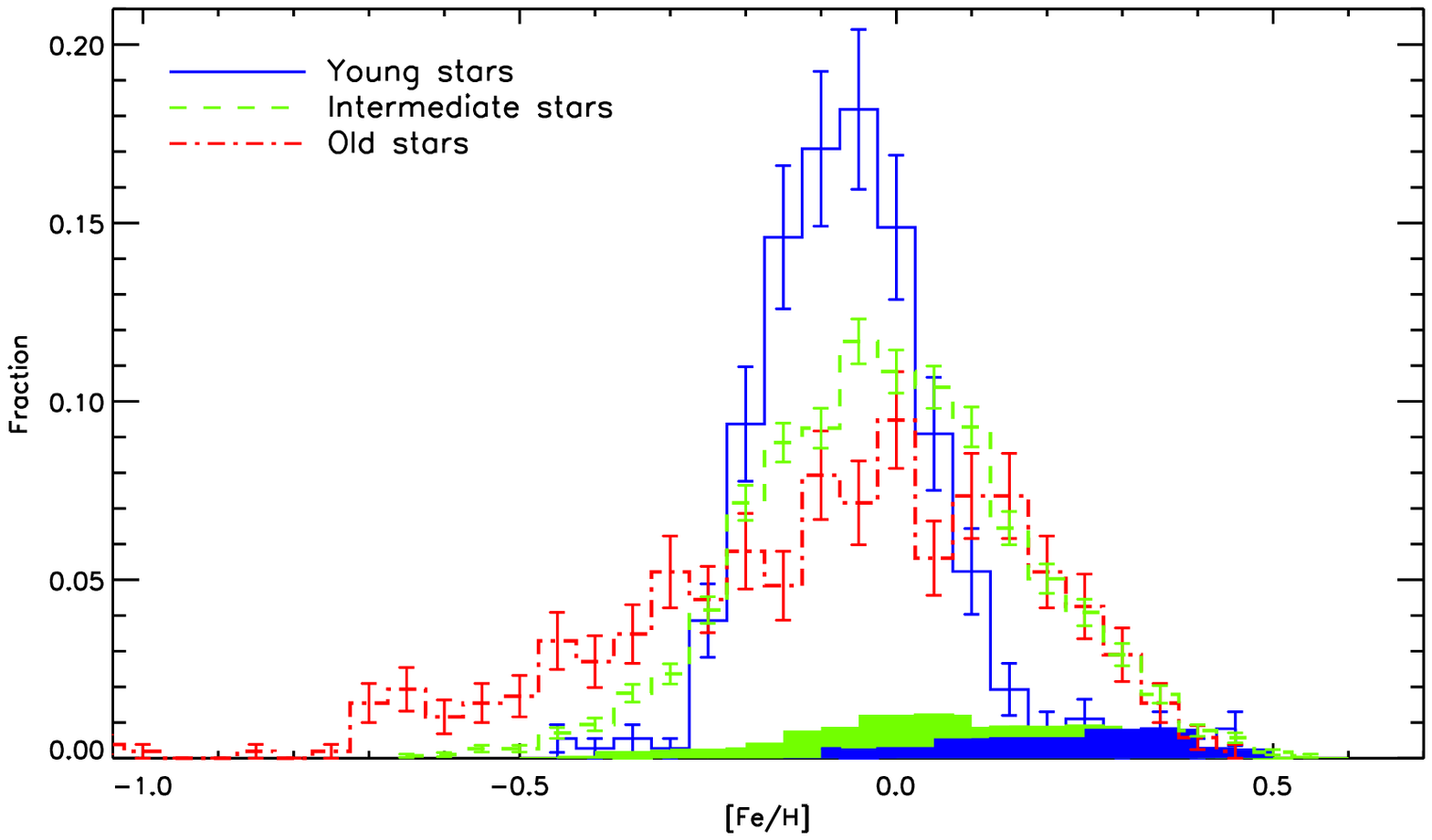 ,width=0.9\hsize}
\epsfig{file=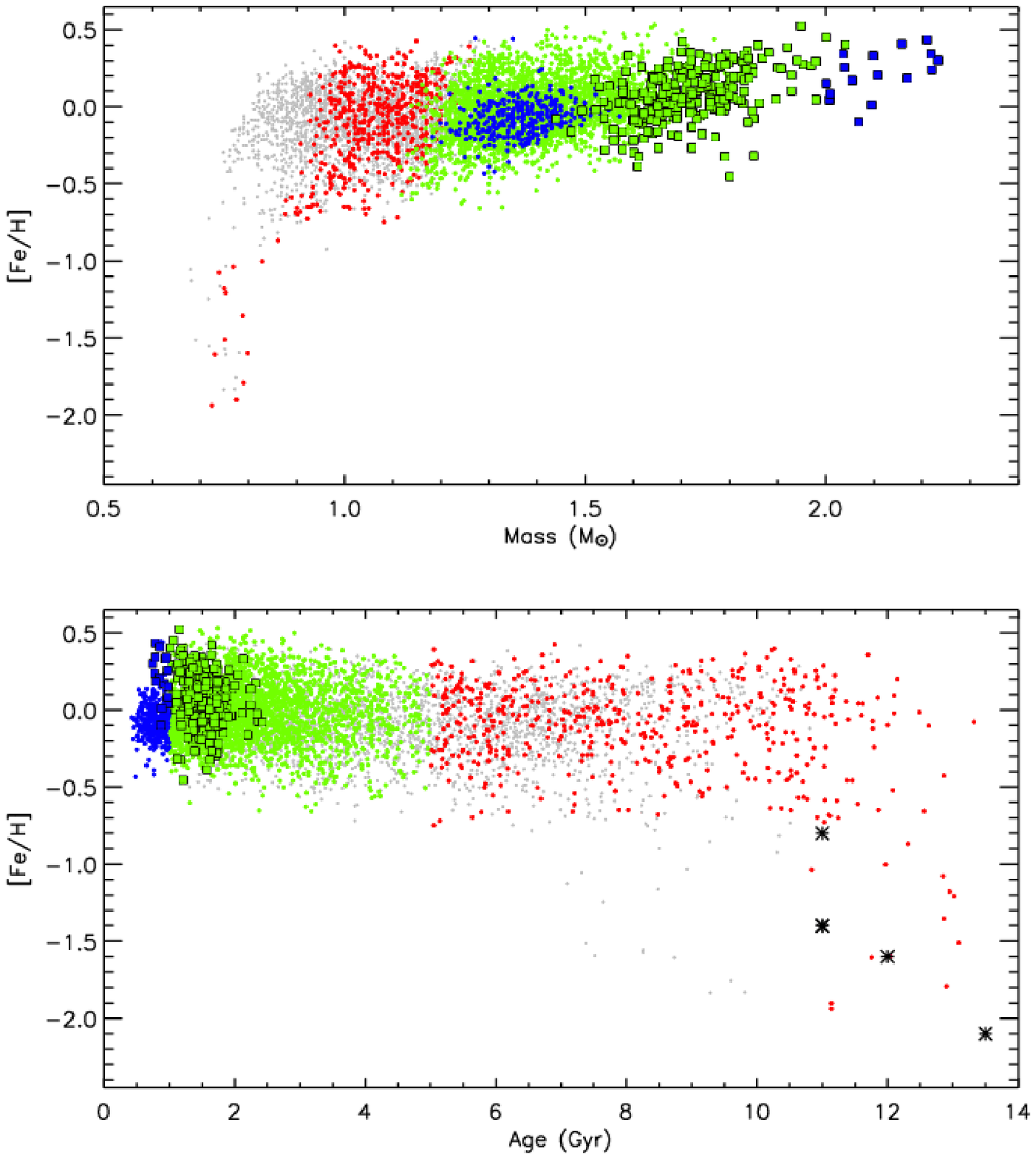,width=0.9\hsize}
\caption{Top panel: MDF for stars belonging to the {\it irfm} sample divided 
into different age intervals. Stars having $\rm{age} < 1$~Gyr are shown with a 
continuous line, $1 \le \rm{age} < 5$~Gyr with a dashed line and $\rm{age} 
\ge 5$~Gyr with a dot-dashed line. Shaded areas identify the subgroup of stars 
in the same age intervals as above, but with absolute magnitudes ($<2$); no 
such bright 
stars are present in the old sample. Only stars with well determined ages 
(see Section \ref{sec:age}) are used. Bars indicate Poisson errors. 
Middle panel: $\feh$ versus stellar mass. Colours have the same meaning 
as in the top panel, with grey dots now referring to the remaining stars having 
more uncertain ages. Filled squares identify stars with bright absolute 
magnitudes ($<2$). Lower panel: same symbols and colours as in the middle 
panel, but showing the age--metallicity relation. Shown for comparison 
(asterisks) are the ages and metallicities of the halo Globular Clusters 
studied in 
\citealt{vandenberg10} (in the latter case, a different zeropoint on the age 
scale is possible, also depending on the input physics adopted in the stellar 
models employed).} 
\label{f:mdfAGE}
\end{figure}

We already argued in Section \ref{sec:intro} that dividing the
original sample into two groups does not introduce any bias. Stars
with the best photometry show lower dispersion, but the average
properties are robust and are the same for both the {\it irfm} and {\it
  clbr} samples. This is shown in Fig.~\ref{f:mdf} for $\feh$ and $\mh$, 
with the relevant statistical parameters given in
Table \ref{t:mdf}.  A Kolmogorov-Smirnov test between the {\it irfm}
and {\it clbr} samples for $\feh$ and $\mh$ tells that
the probability of both samples being drawn from the same distribution
is below 1 percent, i.e.~not significant.  The reason for this lies in
the broader wings of the {\it clbr} sample, partly because the lower
quality of the latter sample could be responsible for less reliably
determined metallicities that over-populate the wings, and/or older ages
(see below). When restricting the selection to $-0.5 \le \feh \le 0.5$,
the {\it irfm} and {\it clbr} samples are in fact drawn from the same
distribution to a level better than 5 percent, under the null
hypothesis that the two distribution are drawn from the same parent
population. Identical conclusions to the Kolmogorov-Smirnov statistic
are reached using instead the Wilcoxon Rank-Sum test for comparison.
We find that the MDF for young and old stars look considerably
different (see below).  We note that because the {\it clbr} sample
contains a few more cooler stars than the {\it irfm} sample (see Section
\ref{sec:irfm}), the cooler stars being preferentially older and thus with
a broader MDF (see below), this could also be partly responsible for
the different broadening of the wings.

Slicing the MDF into different age intervals shows an interesting
feature: young stars have a considerably narrower distribution than
old stars, though the peak always remains around the solar value
(Fig.~\ref{f:mdfAGE}). Notice that because of the selection effects on
the sample age, an uneven slicing --denser at young ages-- is more
appropriate (cf.~Fig.~\ref{fig:agedist}). While the MDF has been
  historically used to constrain the gas infall rate
  \citep[e.g.][]{lb75,tinsley80,mf89,cmg97}, the increasing
broadening with age suggests that old stars are also a relevant
ingredient in describing the wings of the MDF. A natural explanation
is provided by the radial migration of stars \citep{sb02}. In this picture
the solar neighbourhood is not only assembled from local stars,
following a local age metallicity relation, but also from stars
originating from the inner (more metal-rich) and outer (more
metal-poor) Galactic disc that have migrated to the present position on
different timescales \citep[][]{roskar08,ralph09a}. Because of
  the higher density of stars in the inner disc, migration would
  favour metal-rich stars, which could compensate the metal-poor tail
  typical of local chemical evolution, which would explain the rather
  symmetric shape of the MDF we derived. A more quantitative
  explanation, however, requires modelling of the  chemical evolution.

The presence of a metal-rich tail in Fig.~\ref{f:mdfAGE} could be a signature 
of the Galactic bar \citep[e.g.,][]{grenon99}: such a detection is however very 
difficult to claim even with the current sample. 
Indeed, we only detect a conspicuous young metal-rich population at the 
brightest magnitudes, where the accuracy of the metallicity calibration could 
be lower (see the discussion in Section \ref{sec:feh}). The presence of a 
bar would rather imply the existence of an old metal-rich population, which 
we do not detect \citep[but see][for a recent discussion on the effect of the bar]{minchev10,minchev11}. Although we do not have access to the sample selection 
performed in the original assembly of the GCS, we regard the presence of a 
bias against old metal-rich stars as unlikely, and we refer to 
\cite{nordstrom04} for more details on the completeness of the sample. 
We also investigated whether the metal-rich stars display any conspicuous 
feature in the $UV$ velocity plane and did not find any. 
Notice though that the fraction of these young metal-rich stars in the total 
sample is fairly small and they do not bear considerably on the overall MDF 
of Fig.~\ref{f:mdf}.

Apart from the aforementioned bright stars, the metal-rich wing of the MDF is 
not an artefact caused by the sample selection on colours 
\citep[contrary, e.g.,~to][]{kotoneva02}, because high-metallicity stars are 
present throughout the entire mass range (middle panel in 
Fig.~\ref{f:mdfAGE}). Also, on the metal-poor side there is a clear 
contribution of (nearly) unevolved subdwarfs --for which a determination of 
ages is more uncertain-- with a trend in mass mirroring that already observed 
in luminosity (cf.~Fig.~\ref{f:trends} and \ref{f:agemass}). 

\begin{table}
\centering
\caption{Metallicity distribution function}
\label{t:mdf}
\begin{tabular}{lcc}
\hline\hline
	         &       $\feh$      &       $\mh$       \\     
                 &         dex       &        dex        \\
\hline
mean             & $-0.06$ / $-0.07$ & $-0.02$ / $-0.04$ \\
median           & $-0.05$ / $-0.06$ & $-0.01$ / $-0.02$ \\
$\sigma$         & $0.22$  / $0.25$  & $0.19$  / $0.21$  \\
$\rm{FWHM}/2$    & $0.19$  / $0.21$  & $0.17$  / $0.19$  \\
\hline
\end{tabular}
\begin{list}{}{}
\item[] 
{\bf Notes.} Statistical peak values of the MDFs of Fig.~\ref{f:mdf} using 
stars in the {\it irfm} and  {\it clbr} sample. Notice that the MDF is 
influenced by a low-metallicity tail. A Gaussian is not its best description. 
Median and FWHM provide different --and formally better-- estimates.
\end{list}
\end{table}

The peak of the MDF is only slightly subsolar (median $\feh \sim -0.05$, 
$\mh \sim -0.01$), in agreement with e.g.,~\cite{haywood01}, \cite{tc05}, 
\cite{luck06} and \cite{fuhrmann08}, but in contrast with other studies, which 
rather favour a peak in the range $-0.2$ to $-0.1$~dex \citep[e.g.,][]
{wg95,rm96,allende04:s4n,nordstrom04,holmberg07}. In most cases the reason 
for this difference stems from the $\teff$ scale we use, which supports 
spectroscopic studies that adopt similar effective temperatures and results in 
higher metallicities. As a side remark, we note that the MDF determined from 
M dwarfs \citep{bon05:M,casagrande08,casagrande08:uppsala} 
is likely to peak around solar metallicity if the recent spectroscopic 
findings of \cite{ja09} are confirmed and photometric determinations for 
those stars are recalibrated accordingly.

The peak at nearly solar metallicity of the local MDF at all ages 
also has implications 
for understanding secular processes associated with disc evolution, 
by investigating {\it whither and whence} the Sun is moving 
\citep[e.g.,][]{wfd96,bh10}, and it is also an important test of the overall 
solar metallicity \citep{asplund09}. 

\begin{figure}
\epsfig{file=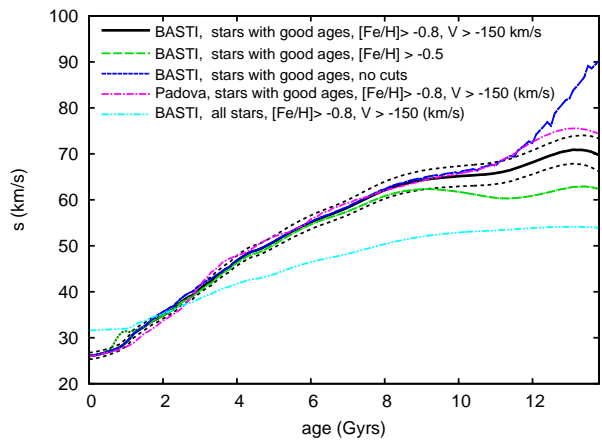,angle=-90,width=0.9\hsize} 
\caption{Velocity dispersion {\tiny $s=\sqrt{\sigma_U^2 + \sigma_V^2 + \sigma_W^2}$} as a function of age. Age probability distribution functions derived from Padova and BASTI isochrones are used. Dotted black lines are $1\,\sigma$ errors for the black line. In all cases stars with $M_{V_T}<2$ are excluded, their effect being responsible for the bump (green dotted line) around $1$~Gyr.
\label{fig:agevelsig}
}
\end{figure}

\section{The age--dispersion relation}\label{sec:adr}

Figure \ref{fig:agevelsig} shows the velocity dispersion $s$ for stars in 
the {\it irfm} sample as a function of stellar age. Ages are determined using 
the BASTI isochrones, apart from one case where the result of using Padova 
isochrones is shown for comparison. The difference between requiring well 
determined ages (according to the definition of Section \ref{sec:age}, black 
line) or not (cyan line) suggests that the signature of a continuous rise 
becomes even more prominent, confirming earlier studies of the GCS 
\citep{nordstrom04,holmberg07}. This has to be expected; because of 
the pronounced overdensity of stars around ages of $2$~Gyr 
(cf.~Fig.~\ref{fig:agedist}), excluding unreliable ages gives less 
contamination to the rarer very young and especially to the older stars. 
However, because velocity dispersion 
roughly increases with age to the power $1/3$ \citep{ss53}, and 
because of the $\sim\,\rm{Gyr}$ uncertainty in ages, it is actually 
difficult to distinguish between a plateau and a real increase.
\begin{figure}
\includegraphics[scale=0.59]{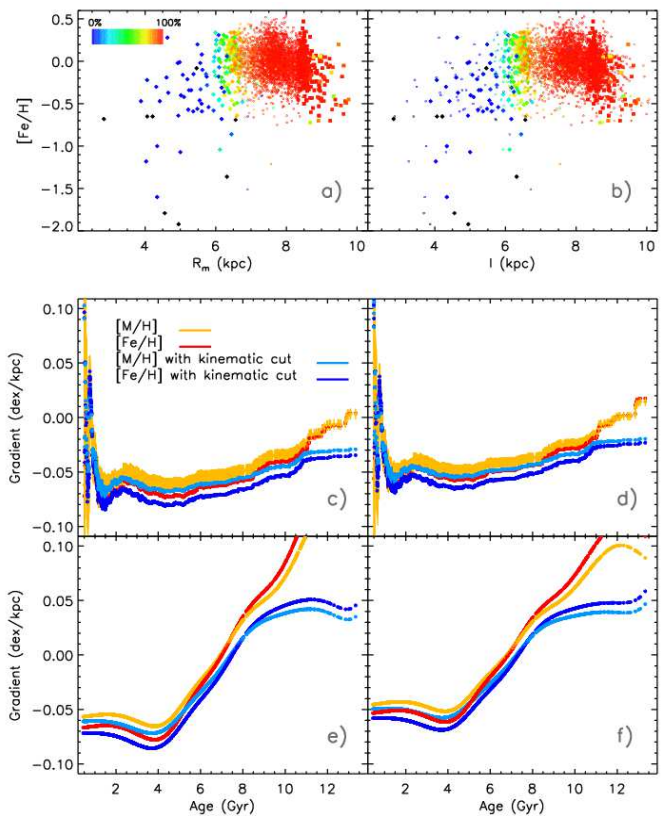}
\caption{Panel a) and b): metallicity as function of orbital and 
guiding centre radius, respectively. The probability of a star
to belong to the thin disc is represented by colour. Stars with halo 
membership higher than 50\% are plotted in black. Filled diamonds and 
squares identify stars having $V_{\rm{LSR}}<-40\,\kms$ and 
$>20\,\kms$, respectively (cf.~with Fig.~\ref{f:velrot}).
Panel c): cumulative metallicity gradient ($\feh$ and 
$\mh$) when including stars of increasingly older ages. Error bars 
are shown in one representative case. The gradient is computed using 
the mean orbital radius of stars as baseline. A kinematic cut to 
exclude halo stars (as described in the text) is also adopted for 
comparison.
Panel d): same panel c), but using the guiding centre radius as baseline.
Panel e): metallicity gradient centred at different ages, weighting 
all other stars with a Gaussian of width $1.5$~Gyr and using the mean 
orbital radius as baseline.  Panel f): same as panel e), but using 
the guiding centre radius. Only stars with well determined ages (Section 
\ref{sec:age}) are used in 
all instances.} 
\label{f:ag}
\end{figure}

When no metallicity nor kinematic cut is applied, a strong rise
appears at the oldest ages (blue line). This feature is likely
 caused by contamination of moderately metal-poor stars that
might belong to the Galactic halo. This disappears
when using a very conservative cut at $\feh>-0.5$~dex or a milder one
at $\feh>-0.8$~dex but only considering stars with $V > -150\,\kms$.
These cuts exclude some tens of stars, consistent with
  expectations from local disc-to-halo normalization which, despite
  large uncertainties, is in the range of a few hundreds-to-one
  \citep[e.g.][]{morrison93,gfb98,juric08} Because the isochrones
might fail to exactly match metal-poor stars (Section \ref{sec:age}),
the derived age distribution of low metallicity stars can be biased to
older ages. Difficulties in understanding selection criteria are 
likely to be responsible for the different findings of \cite{qg00} who 
--essentially using the sample of 189 stars studied in \cite{edvardsson93}-- 
claimed the presence of a plateau in the dispersion over all intermediate ages
followed by a quick rise at about $\sim 10$~Gyr.

\section{Disc}\label{sec:disc}

\subsection{Metallicity gradient}

Abundance gradients across the Galactic stellar disc provide fundamental 
constraints on the chemical evolution of this component of the Milky Way, and 
on the physical assumptions adopted in chemical evolution models 
\citep[e.g.,][]{pc99,chia01}. Despite its local nature, the large number of 
stars in the GCS would suggest that it is possible to use it for estimating the 
radial metallicity gradient in the Galaxy \citep[cf.~e.g.,][]{nordstrom04}.

While Galactic radial positions ($R_{\rm{Gal}}$) are snapshots of 
stars at the present time, covering a very limited range in distances 
(at most $0.3-0.4$~kpc for the GCS), their mean orbital radii 
$R_{\rm{m}}$ (left 
panels in Fig.~\ref{f:ag}) allow us to probe larger distances (up to a 
few kpc) and thus are better suited for deriving the metallicity 
gradient \citep[e.g.][]{nordstrom04,holmberg07}. Orbital radii 
depend on the adopted Galactic potential; $R_{\rm{Gal}}$ and $V$ 
velocities offer an alternative and model-independent approach via 
the guiding centre radius $I=\frac{R_{\rm{Gal}}(V+232)}{220}$ under the 
assumption of a constant circular rotation speed of $220\,\kms$ 
(right panels in Fig.~\ref{f:ag}).

Using only stars that belong to the {\it irfm} sample and are within the
calibration range, the exact value of the gradient still depends on
whether or not a cut at the lowest metallicities is imposed to exclude
contamination from halo stars. In Fig.~\ref{f:ag} we show the
  case of applying neither kinematic nor metallicity cuts, as well as
a kinematic selection to retain only stars with probability higher
than $90$ percent of belonging to the thin or thick disc \citep[see
e.g.][for more details on this kind of selection
procedure]{ramirez07}. Cutting the sample to exclude
  metallicities lower than about $-0.8$~dex has a similar effect as
  the kinematic selection. Metal poor (halo) stars having small
  orbital (and guiding centre) radii are in fact responsible for the
  strong positive rise in the metallicity gradient.

It appears obvious that taking all stars at their face values 
does not provide a meaningful measure of the gradient in the disc. 
Indeed its value 
depends on the adopted kinematic or metallicity cuts, the age 
interval considered, and also whether orbital or guiding centre
radii are used for the computation 
(Fig.~\ref{f:ag}). Difficulties in estimating e.g. the interdependence 
between age and kinematic cuts (as stars with increasing asymmetric 
drift are preferentially older) as well as the increasing scatter in 
the age--metallicity relation and in the age--dispersion 
relation further complicate the picture.

Fig.~\ref{f:ag} (middle panels) suggests the presence of a moderate 
negative radial 
gradient, consistent with studies using other indicators at various 
Galactocentric distances such as Cepheids, HII regions, B stars, 
open clusters and planetary nebulae 
\citep[see e.g.,][and references therein]{mc10_gra}. When restricting 
the analysis to different age intervals (lower panels) there is an 
indication of a flattening and even a reversal of the gradient with 
increasing age, but we stress once more that the adopted kinematic 
or metallicity cuts affect the results. Such a signature comes from 
older thick (as well as halo) stars \citep{spagna10}, while the 
GCS is mostly limited to the younger objects that are situated in 
the thin rather than the thick disc. 

As already mentioned in Section \ref{sec:mdf}, the different behaviour for 
younger and older stars can be understood in terms of radial migration, where 
the increasing age that is responsible for a broadening of the MDF could also 
soften the gradient, but more data and extended analyses are needed to 
explore this scenario. 

\subsection{Thin, thick or stirred?}\label{sec:tts}

Observations of external edge-on galaxies show the presence of both a thin and 
thick disc component \citep{burstein79,db02,yd06}. The Milky Way seems to have 
a two-component disc as well, which was first proposed to fit the vertical 
density profile derived from star counts \citep{yo82,gr83}. Disentangling the 
nature and origin of these components is therefore highly relevant for 
understanding galaxy formation.
While models in which thick and thin discs form sequentially via a rapid or 
dissipative collapse of protogalactic clouds became disfavoured during the 
past years \citep[e.g.,][]{m93}, it is not yet clear how the stellar disc can 
form a thick component with time, if this is caused by to internal \citep[scattering, dynamical interaction or radial mixing, e.g.,][]{ralph09b,loe10} or external \cite[satellite accretion, mergers of gas-rich systems, minor mergers, e.g.,][]{abadi03,brook07,villa08,scanna09} mechanisms.

Though limited to the solar neighbourhood, the GCS can provide important 
insights into this puzzle, because it is essentially free from kinematic 
selections. Our metallicities and 
$\alfe$ (Section \ref{sec:alpha}) provide for the first time a way 
to investigate this with a more complete sample. 

\begin{figure}
\includegraphics[scale=0.72]{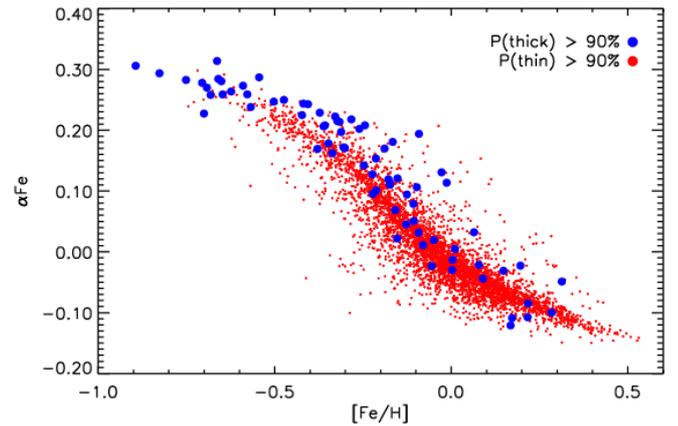}
\caption{$\feh$ vs.~$\alfe$ for stars in the {\it irfm} sample within the 
metallicity calibration ranges and with kinematic information to assign 
statistical 
membership to the thin or thick disc. Only stars with a membership probability 
higher than 90 percent are shown (4655 stars in total). A Gaussian noise of 
$0.005$~dex was added on both axes for better displaying all stars.} 
\label{f:tt}
\end{figure}

Figure \ref{f:tt} shows all stars with a reliably determined $\feh$ and 
$\alfe$ and for which the $U,V,W$ velocities are known, so that the same 
kinematic 
probabilistic selection scheme to the thin or thick disc adopted in the 
previous section can be applied \citep{ramirez07}. 
The small scatter and overall shape of the plot simply reflects the fiducial 
used to derive $\alfe$, which squeezes up most of the metal-poor 
stars and also prevents us from seeing any gap between the thin and the thick 
discs (see discussion in Section \ref{sec:alpha}). Despite these limitations, a 
qualitative picture can be drawn. Stars kinematically attributed to the thick 
disc populate the upper envelope of the Fig.~\ref{f:tt} for subsolar 
metallicities, while merging into the thin disc around solar $\feh$, in 
agreement with similar findings obtained by studies based on high-resolution 
spectroscopy \citep[e.g.,][]{reddy06,bensby07}. 

\begin{figure*}
\includegraphics[scale=0.87]{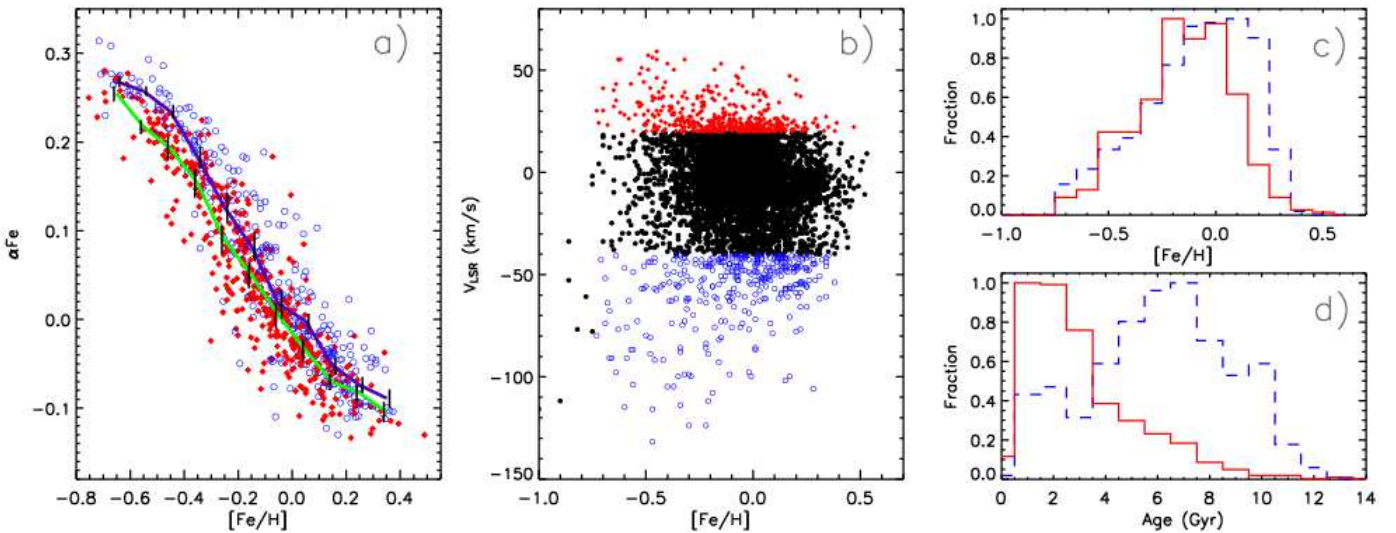}
\caption{Panel a): $\alfe$ for stars in the {\it irfm} sample having 
$V_{\rm{LSR}}=V+V_{\odot}>20\,\kms$ (filled diamonds) or $<-40\,\kms$ (open 
circles). Only $\feh > -0.73$~dex were selected to avoid a metal-poor 
tail in stars with negative velocities (the cut in $\feh$ was selected to 
be the same as the lower-most value encountered in stars having 
$V_{\rm{LSR}}>20\,\kms$, but its exact choice is anyway irrelevant for the 
discussion). 
A few stars with clear halo kinematics were also excluded. Lower/upper 
continuous line connects the mean $\alfe$ in different 
$\feh$ intervals for filled diamonds/open circles. Error bars (slightly 
offset in abscissa for clearer comparison) are the standard deviation of the 
mean in each 
$\feh$ bin. Panel b): $V_{\rm{LSR}}$ as function of metallicity for all 
stars in sample {\it irfm} with kinematic information. Filled 
diamonds and open circles as in the previous panel. Panel c) and d): 
normalized metallicity and age distributions for the two previous group of 
stars 
having $V_{\rm{LSR}}>20\,\kms$ (continuous line) or $<-40\,\kms$ (dashed line). 
The value $V_{\odot}=12.24\;\kms$ was adopted \citep{sbd10}.} 
\label{f:velrot}
\end{figure*}

Similarly, Fig.~\ref{f:velrot} shows the $\alfe$ vs.~$\feh$ plane for stars 
belonging to the {\it irfm} sample. Stars are separated by 
their rotation velocities ($V$) as depicted in the middle panel showing 
$V$ vs.~$\feh$. Clearly, this is only a rough criterion for the division and 
this selection is not stringent in targeting single disc ``components'', yet a 
striking difference appears. At each metallicity, the stars with high
negative $V$ velocities (open circles) have higher average $\alfe$; the 
difference is indeed small in terms of $\alfe$, but statistical significant. 
This can be expected because stars with such a large asymmetric drift 
should be significantly older than the remaining population (because of the 
asymmetric drift--dispersion and the age--dispersion 
relations), which is  confirmed by their age distribution in the lower right 
panel, which is indeed far older. Our analysis thus clearly confirms 
a similar result drawn by \cite{haywood08} from a smaller 
spectroscopic sample. Comparison with Fig.~\ref{f:ag} also shows the 
correspondence between our identification based on rotational 
velocities and the orbital radii of stars. It is interesting to 
notice that \cite{edvardsson93} found a hint that stars with high/low orbital 
radii lie on the lower/upper envelope of the $\aFe$ vs.~$\feh$ plot, 
consistent with what we see here.

Because of the tight age--metallicity relation in chemical evolution
models without radial migration \citep[e.g.,][]{cmg97}, older and
alpha-richer stars are expected to be more metal-poor. Yet
Fig.~\ref{f:velrot} rather tells the opposite, with the old and
alpha-rich stars also being on average more metal-rich than the
population with high rotation velocities (filled diamonds). The
emergence of a metal-rich, old thick disc was already present in
spectroscopic sample of \cite{fb08}. This apparently surprising
behaviour is however readily explained if there is no strong
age--metallicity dependence, as is the case in radial migration
models, and if the lagging metal-rich population comprises --to some
extent at least-- objects from the inner disc, which are more
metal-rich thanks to the Galactic metallicity gradient
\citep{ralph09a}.

\section{Conclusions}\label{conclusions}

Low mass, long lived stars are crucial witnesses of the chemical and dynamical evolution of the Milky Way, but to properly harvest this information, we must ensure that we have determined their astrophysical parameters to the highest accuracy possible, given the observational limitations. The Geneva-Copenhagen Survey provides the ideal database to achieve this goal: it is kinematically unbiased, all its stars have highly homogeneous Str\"omgren photometry, from which stellar abundance information can be readily derived and merging this catalogue with Tycho2 and 2MASS provides the multi-band optical and infrared photometry needed to derive $\teff$ via the infrared flux method. 

We have carried out a revision of the GCS not only benefiting from the latest developments in setting the zeropoint of the effective temperature scale, but also improving upon the homogeneity of the stellar parameters for all stars in the sample. In comparison to previous GCS calibrations, our effective temperatures are hotter; at the same time the improved methodology often reduces the intrinsic uncertainty per star to below $100$~K. This leads to a much better agreement between stars and isochrones in the HR diagram, which allows us to directly derive ages via a Bayesian approach. Because we did not make use of metallicity--dependent temperature shifts to reconcile isochrones with data, the risk of introducing an artificial age--metallicity relationship is reduced. Since the adopted effective temperature scale has immediate consequences on abundances, we recalibrated Str\"omgren indices versus stellar metallicities using a sample of nearly $1500$ stars with high-resolution spectroscopic abundances derived adopting $\teff$ consistent with ours. We thus warn that when comparing our results with other studies, it should always be kept in mind that differences in metallicities could simply reflect the different $\teff$ scales adopted.
As a consequence, the mean metallicity of previous GCS analyses is increased by $\sim 0.1$~dex, now peaking at $\mh \sim -0.01$~dex and thus making the Sun a completely average star given its metallicity \citep[see also][]{asplund09}. 
It is intriguing to note that in the past the higher metallicity of 
the Sun compared to local dwarfs was used in support of radial 
migration \citep[e.g.][]{wfd96}; our analysis suggests that 
the Sun is not atypical, at least in metallicity. Instead, we 
derive other atypical properties for disc stars, to explain which, 
radial migration could be a relevant ingredient.

For the first time we are able to derive $\aFe$ estimates from Str\"omgren photometry (named $\alfe$ for the sake of clarity). The method becomes more unreliable for increasingly hotter objects and also for metal-poor stars (roughly below $-1$~dex), but focusing on disc stars gives a reasonable guidance on the {\it relative} alpha enhancements for the whole sample. The ability to reach this tentative distinction enabled us to bring the metallicity calibration to significantly better accuracy by reducing the uncertainty in $\aFe$ enhancement. The new metallicity scale was then checked against open clusters and a moving group, showing indeed a high degree of internal consistency with a suggested intrinsic scatter below $0.10$~dex in $\feh$. The recently measured $uvby$ solar colours finally corroborate the agreement between the temperature and metallicity scales. 

Having this at hand, we revised and complemented the largest existing sample of F and G dwarf stars in the solar neighbourhood that is kinematically unbiased and gives information on ages, the abundance plane, and kinematics. A preliminary analysis of this dataset supports a scenario with a strong interplay among those three characters: the metallicity distribution function shows increasing broadening at older ages, suggesting that its wings could mostly comprise stars born at various Galactocentric radii and migrated at the current position over different timescales. This scenario could also account for the radial gradient getting flatter for older ages, though this detection is yet uncertain partly because of the short distance baseline covered by the GCS, and partly because of the difficulties in disentangling metallicities, ages, and kinematic selection in the sample. 

A more robust and striking feature comes instead from the division of stars in the rotation velocity plane, which are shown to have different patterns in the abundance plane and in ages, a feature which is unexpected in classical chemical evolution models, but seen in spectroscopic studies and naturally explained if stellar radial migrations is taken into account. Despite that our data show clear support for the radial migration scenario, many different processes enter the picture of galaxy formation and evolution; future larger surveys will thus be invaluable to further constrain the interplay of various scenarios. The results presented here are thus an example of the importance of having at the same time kinematic, metallicity, and age information to uncover the past of our Galaxy.

\begin{acknowledgements}
  It is a pleasure to acknowledge an anonymous referee for his/her
  insightful comments and excellent review, which has considerably
  strengthened the paper. We thank Chris Flynn and Laura Portinari for a
  careful reading of the manuscript and useful discussions. We are
  indebted to Birgitta Nordstr\"om, Johannes Andersen and Johan
  Holmberg for useful discussions on this revision, providing
  Str\"omgren colours in advance of publication and for previous
  versions of the Geneva-Copenhagen Survey, without which our work
  would have not been possible in the first place. This work was performed in 
  part (IR) under contract with the California Institute of Technology 
  (Caltech) funded by NASA through the Sagan Fellowship Program. SF's research 
  is partly supported by the grant 624-2008-4095 from the Swedish Research
  Council. This publication makes use of data products from the Two
  Micron All Sky Survey, which is a joint project of the University of
  Massachusetts and the Infrared Processing and Analysis
  Center/California Institute of Technology, funded by the National
  Aeronautics and Space Administration and the National Science
  Foundation.
\end{acknowledgements}

\bibliographystyle{aa}
\bibliography{refs}

\begin{appendix}

\section{Bayesian age determination}

For the effective temperatures we assume a Gaussian error, which is derived for each star as described in Section \ref{sec:irfm} or \ref{sec:clbr} for the 
{\it irfm} and {\it clbr} sample, respectively.
Things get slightly more complicated for the magnitude errors. The magnitude is estimated from the photometric measurements and the parallax of these stars, while the latter measurement completely dominates the error. So, assuming a Gaussian distribution in the parallax, we can write
\begin{equation}
P_p(p | p_0, \sigma_p) = \frac{1}{\sqrt{2\pi}\sigma_p}e^{\frac{-(p-p_0)^2}{2\sigma_p^2}},
\end{equation}

where $p_0$ is the best estimate for the parallax, $\sigma_o$ the adopted parallax error. Converting to magnitude space we thus have
\begin{equation}
\begin{array}{ll}
P_V(p(V) | V_0, \sigma_p) & = P_p(V(p) | V_0, \sigma_p) \frac{dp}{dV} = \\ &\\
& =  k10^{0.2 \Delta V} \exp\left({-\frac{{\left(10^{0.2 \Delta V} - 1\right)}^2}{2 \sigma_{pr}^2}}\right),
\end{array}
\end{equation}
where $\Delta V = V - V_0$ is the difference between the magnitude $V$ and the best parallax-based estimate for the magnitude $V_0$, $\sigma_{pr} = \sigma_p / p_0$ is the relative parallax error and $k$ is some normalization constant. For small relative parallax errors this treatment does not imply any significant changes, because the error distribution approaches a Gaussian. However, for $\sigma_{pr} \gtrsim 0.1$, the $V$ magnitude distribution becomes increasingly skewed, because the lower parallaxes produce an extended tail towards brighter magnitudes \citep[see also][for an analytic estimate of the bias in case of low parallax errors]{casagrande07}.

It is crucial to correct the estimated metallicities for systematic biases. In the wings of the MDF more stars are scattered out from the more densely populated central regions, so that there is a net bias to be expected in the metallicities in the wings of the distribution, e.g. stars on the high-metallicity wing have on average overestimated metallicities, while stars on the left wing of the MDF are expected to have an increased fraction of metallicity underestimates. Because of these shifts, a na\"ive use of the measured metallicities would introduce an age underestimate on the high-metallicity side and an age overestimate on the low-metallicity wing, which would give rise to an artificial age--metallicity correlation. 
This bias can be reduced and in the best case removed by an appropriate metallicity prior that reflects the underlying ``real'' metallicity distribution.
It might be tempting to use the metallicity distribution itself as metallicity prior in an iterative process. However, this is intrinsically unstable, because stars would assemble in peaks, growing by attraction of more objects. So we took an analytical function that approximates the sample distribution:
\begin{displaymath}
f(\mh) = 
\end{displaymath}
\begin{displaymath}
= \left\{
\begin{array}{lll}
{387.8 m(\mh)}&for& \mh \ge 0.04\\ &&\\387.0 m(\mh) cor(\mh) + 0.8  &for& \mh < 0.04 
\end{array}
\right.
\end{displaymath}
with
\begin{displaymath}
m(\mh) = \exp\left({-\frac{(\mh - 0.04)^2}{2 \cdot 0.12^2}}\right)
\end{displaymath}
\begin{displaymath}
cor(\mh) = 1 + 0.3 (e^{-20(\mh + 0.26)} - e^{-6.0}).
\end{displaymath}
On the right hand side we simply choose a Gaussian term as prior. On the left hand side this is considerably flattened by adding the ``correction'' term. 
This function has to be multiplied with the Gaussian error term in metallicity. Mainly its relative slope decides about shifts in the adopted probability distribution in metallicity. So with the correction term that flattens the distribution at low metallicities we can hope to reproduce the actual data sufficiently 
well.
\begin{figure}
\includegraphics[scale=0.88]{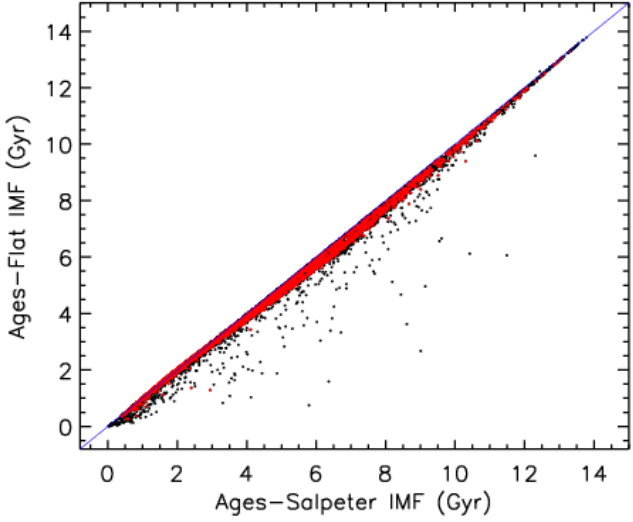}
\caption{Comparison between ages derived assuming the Salpeter vs.~flat IMF. 
Red dots are stars belonging to the {\it irfm} sample.} 
\label{f:imf}
\end{figure}

\begin{figure}
\includegraphics[scale=0.88]{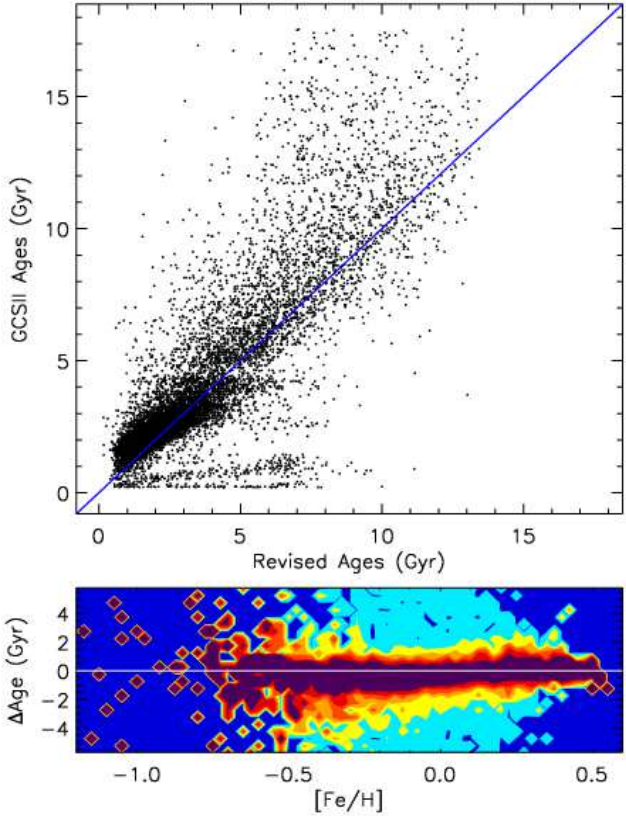}
\caption{Comparison between our revised ages on those in GCSII for the full 
catalogue. $\Delta\rm{Age}$ is in the sense ours minus GCSII. Contour levels in 
the lower panel same as in Fig.~\ref{f:teff}.} 
\label{f:ageplot}
\end{figure}

Altogether the probability distributions in each parameter of a star are gained by running over the isochrone points.
The probability distribution in one parameter $x_i$ is gained by
\begin{displaymath}
\begin{array}{ll} 
P(x_i) = & k_P \sum_{is}{\delta(x_i - x_{i, is}) {\mathcal A}({\tau}_{is}) IMF(m_{in, is})}\\ &\\
& \cdot P_V(V_{is} | V, \sigma_p)G(log(\teff)-log(T_{\rm{eff}, is}), \sigma_T) \\ & \\
& \cdot f(\mh_{is})G(\mh - \mh_{is}, \sigma_{\mh}) \\ & \\
& \cdot  d_{m_{in}, is} d_{\tau, {is}} d_{\mh, is}\,,
\end{array}
\end{displaymath}
where the sum runs over all isochrone points $is$. $G((y-y_0), \sigma_y)$ is a Gaussian function with $y-y_0$ in the counter of the exponent and with dispersion $\sigma_y$, $m_{in, is}$ is the initial mass of the star on an isochrone point, ${\mathcal A}$  is the age prior, $d_m$, $d_{\tau}$ and $d_{\mh}$ denote the effective volume covered by an isochrone point, i.e. half the distances to its neighbours in initial mass, age and $\mh$, and subscript $is$ denotes the values of the isochrone point. From the probability distribution the maximum likelihood, median, expectation and $5$, $16$, $84$, $95$ percent values of the age are derived. The error $\sigma$ for any given age is defined as half difference of the $84$ minus $16$ percent value and the relative uncertainty as the ratio of 
$\sigma$ over expectation age. Throughout the paper, BASTI expectation ages and masses are used in the analysis if not otherwise specified. 
In all instances a Salpeter IMF was used, and the effect 
of this choice on $P(x_i)$ is minimal. Fig.~\ref{f:imf} shows the 
difference in BASTI expectation ages when a flat IMF is used 
instead.

A comparison between our BASTI expectation ages and those derived in GCSII is 
shown in Fig.~\ref{f:ageplot}. Age determinations are subject to many 
subtleties: because we do not have access to some technicalities used in 
GCSII, it is difficult to explain all trends that arise in the comparison. 
Some of the breaks (e.g.~the depletion of stars around 5~Gyr) could arise 
because isochrones in GCSII were shifted to agree with the data. 
Similarly, we notice that stars in GCSII with undetermined lower 
confidence limits are preferentially assigned young ages, 
which are responsible for some of the horizontal stripes seen in 
the upper panel of Fig.~\ref{f:ageplot}.

\end{appendix}

\end{document}